# The interactions of SARS-CoV-2 with co-circulating pathogens: Epidemiological implications and current knowledge gaps

Anabelle Wong[1,2,*], Laura Andrea Barrero Guevara[1,2,*], Elizabeth Goult[1,*], Michael Briga[1], Sarah C. Kramer[1], Aleksandra Kovacevic[3,4], Lulla Opatowski[3,4], Matthieu Domenech de Cellès[1,**]

1. Infectious Disease Epidemiology group, Max Planck Institute for Infection Biology, Charitéplatz 1, 10117 Berlin, Germany.
2. Institute of Public Health, Charité–Universitätsmedizin Berlin, Charitéplatz 1, 10117 Berlin, Germany.
3. Epidemiology and Modelling of Antibiotic Evasion, Institut Pasteur, Université Paris Cité, Paris, France
4. Anti-infective Evasion and Pharmacoepidemiology Team, CESP, Université Paris-Saclay, Université de Versailles Saint-Quentin-en-Yvelines, INSERM U1018 Montigny-le-Bretonneux, France.

* These authors contributed equally.
** Corresponding author: Dr. Matthieu Domenech de Cellès, Max Planck Institute for Infection Biology, Charitéplatz 1, Campus Charité Mitte, 10117 Berlin, Germany

**Abstract**

Despite the availability of effective vaccines, the persistence of SARS-CoV-2 suggests that co-circulation with other pathogens and resulting multi-epidemics (of, for example, COVID-19 and influenza) may become increasingly frequent. To better forecast and control the risk of such multi-epidemics, it is essential to elucidate the potential interactions of SARS-CoV-2 with other pathogens; these interactions, however, remain poorly defined. Here, we aimed to review the current body of evidence about SARS-CoV-2 interactions. To study pathogen interactions in a systematic way, we first developed a general framework to capture their major components: sign (either negative for antagonistic interactions or positive for synergistic interactions), strength (i.e., magnitude of the interaction), symmetry (describing whether the interaction depends on the order of infection of interacting pathogens), duration (describing whether the interaction is short-lived or long-lived), and mechanism (e.g., whether interaction modifies susceptibility to infection, transmissibility of infection, or severity of disease). We then reviewed the experimental evidence from animal models about SARS-CoV-2 interactions. Of the fourteen studies identified, eleven focused on the outcomes of co-infection with non-attenuated influenza A viruses (IAV), and three with other pathogens. The eleven studies on IAV used different designs and animal models (ferrets, hamsters, and mice), but generally demonstrated that co-infection increased disease severity compared with either mono-infection. By contrast, the effect of co- infection on the viral load of either virus was variable and inconsistent across studies. Next, we reviewed the epidemiological evidence about SARS-CoV-2 interactions in human populations. Although numerous studies were identified, only few were specifically designed to infer interaction and many were prone to multiple biases, including confounding. Nevertheless, their results suggested that influenza and pneumococcal conjugate vaccinations were associated with reduced riskof SARS-CoV-2 infection. Finally, we formulated simple transmission models of SARS-CoV-2 co-circulation with a viral or a bacterial pathogen, showing how they can naturally incorporate the proposed framework. More generally, we argue that such models, when designed with an integrative and multidisciplinary perspective, will be invaluable tools to resolve the substantial uncertainties that remain about SARS-CoV-2 interactions.

# 1. Introduction

As of August 2022, the pandemic of coronavirus disease 2019 (COVID-19)—caused by the novel severe acute respiratory syndrome coronavirus 2 (SARS-CoV-2)—has resulted in at least 598 million cases and 6.4 million deaths worldwide [1]. Despite the implementation of stringent control measures and the increasing roll-out of effective vaccines in many locations, the persistent circulation of SARS-CoV-2 suggests the infeasibility of elimination and the gradual transition to endemic or seasonal epidemic dynamics [2]. Hence, co-circulation of SARS-CoV-2 with other pathogens may become increasingly frequent and cause multiple simultaneous epidemics of, for example, COVID-19 and influenza [3].

Interaction—that is, the ability of one pathogen to alter the risk of infection or disease caused by another pathogen (Fig. 1)—is an essential aspect to forecast the dynamics of co-circulating infectious diseases. From a public health perspective, interactions may significantly aggravate disease burden, as demonstrated for immunosuppressive viruses like measles [4] and human immunodeficiency virus (HIV) [5]. Another interesting, yet understudied public health implication of interactions is the possibility of indirect effects of vaccines on non-target pathogens, as suggested for influenza vaccines [6,7] . However, despite their potentially large relevance to SARS-CoV-2 epidemiology and COVID-19 control measures, the interactions of SARS-CoV-2 with other pathogens remain poorly defined.

Here, we aimed to review the current body of evidence about the interactions of SARS-CoV-2 with co-circulating pathogens. We first present a general framework to capture the complexities of interactions and study them in a systematic way. Using this framework, we then review the results of published experimental and epidemiological studies. Finally, we formulate simple transmission models incorporating the proposed framework to illustrate the potential population-level impact of SARS-CoV-2 interactions.

## 2. Dissecting pathogen interactions: Sign and Strength, Timing, and mechanisms

Pathogen interactions can be complex, because of the multiple elements needed to fully characterize them. To study interactions in a systematic and comprehensive way, we propose a conceptual framework—depicted schematically in Fig. 1—that incorporates three essential components of interaction, detailed below.

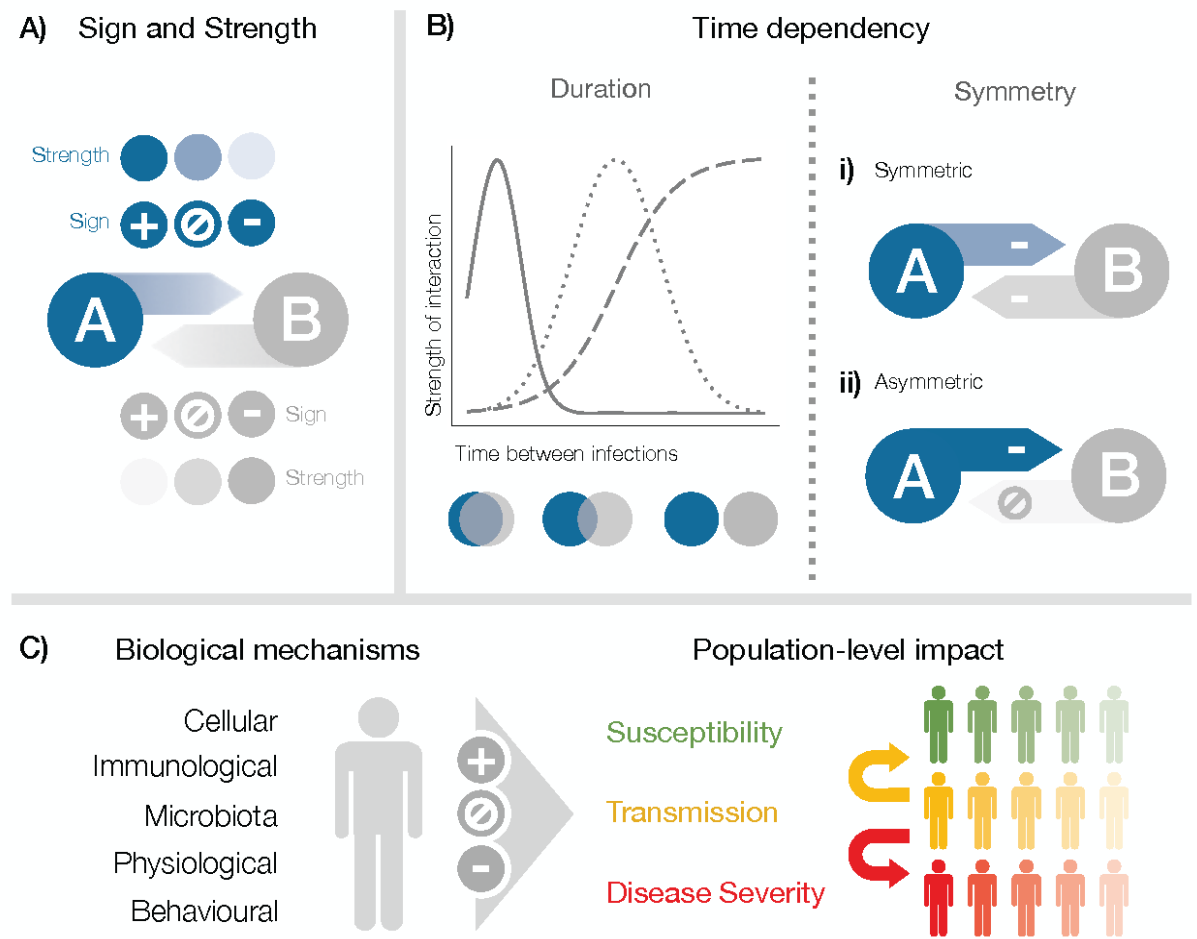

**Figure 1. A conceptual framework to study pathogen interactions**. For a given pair of pathogens, interaction can be characterized by its sign and strength (A), which in turn depend

on the time interval between infections (duration of interaction) and on the sequence of infection (symmetry of interaction) (B). Examples include negative, symmetric interactions (as in the case of influenza B virus Victoria lineage and Yamagata lineage) and negative, asymmetric interactions (as in the case of influenza A virus and respiratory syncytial virus). Interaction can be caused by different biological mechanisms (C), which determine its positive or negative effects on susceptibility to infection, transmission (transmissibility and duration of infection), or disease severity at the individual level and in turn its impact at the population level.

**2.1 Sign and strength of interaction**

The first dimension of this framework is the sign and strength of interaction. Here, we define the sign of interaction as positive in synergistic interactions (where a first pathogen increases the risk of infection or disease of a second pathogen) and negative in antagonistic interactions (where the risk is decreased), and we refer to strength as the magnitude of effect on a given parameter exerted by one pathogen on another.

An example of negative interaction exists between influenza A virus (IAV) and human respiratory syncytial virus (RSV), for which experimental studies have shown that a recent IAV infection inhibits the growth of RSV in ferrets [8] and in mice [9]. By contrast, IAV interacts positively with *Streptococcus pneumoniae* (the pneumococcus) by promoting bacterial growth [10,11]. This illustrates that interaction is pathogen-specific and cannot be easily extrapolated to other pathogen systems.

**2.2 Time-dependency of interaction**

The second dimension of our proposed framework is time-dependency: both the time between infections and the sequence of infection can affect the sign and strength of an interaction.

Duration of interaction and time between infections

Due to the kinetics of cellular and humoral immune response following respiratory infections [12–14], the strength of interaction can change with time between infections. For example, primary IAV infection prevented subsequent RSV infection in ferrets when exposed 3 days later but the protection disappeared as the time between IAV and RSV challenges increased to 11 days [8]. Such short-lived negative interaction was also observed between influenza B virus Victoria lineage (B/Vic) and Yamagata lineage (B/Yam) [15]. Interaction can be long-lived if it is mediated by immune memory. For example, measles infection can partially erase previously acquired immunity to other pathogens, causing "immune amnesia" [16]. Childhood exposures to a given IAV subtype can cause long-lasting immunological bias that shapes the individual's subsequent risk for influenza infection [17].

Symmetry of interaction and sequence of infection

The sequence of infection can also affect the interaction, as evidenced by the asymmetric effects found in previous studies. For example, prior infection with IAV or RSV hindered rhinovirus (RV) replication, but prior RV infection did not interfere with IAV and RSV replication in human airway epithelium [18]. While IAV infection predisposed individuals to pneumococcal colonization and infection [19–21] and led to more severe disease [22], evidence from animal and human challenge studies demonstrated that prior pneumococcal colonization did not lead to more severe disease [20,23,24] but might have had a protective effect against viral replication [24,25] upon subsequent IAV challenge. Interestingly, this effect might depend on the density of pneumococcal colonization [20,23,24].

By contrast, when a negative interaction is symmetric between two pathogens, whichever of the two pathogens is the first to infect can inhibit subsequent infection by the other pathogen—as in the case of influenza B lineages [15].

**2.3 Biological mechanisms and population-level impact of interaction**

The third dimension in our framework is the mechanism of interaction: interaction can be caused by different biological mechanisms, which determine its positive or negative effects on susceptibility to infection, characteristics of infection (such as transmissibility and duration), or disease severity at the individual level and in turn its impact at the population level (Fig. 1C).

Biological mechanisms

Examples of biological mechanisms of pathogen interaction include intra-cellular and physiological changes and effects on the immune response, on the respiratory microbiota, and on host behaviors. A pathogen can induce **changes on the host cells** that are beneficial or detrimental to another pathogen. For example, it has been shown that RSV and human parainfluenza virus 3 (HPIV-3) increase expression of receptors for *Haemophilus influenzae* and the pneumococcus binding in bronchial epithelial cells [26]. In both cases, changes in cellular expression may lead to a positive interaction. A pathogen can cause **changes to the host's immune profile** (e.g., depletion of CD4+ T cells by HIV [5], increased IFN response by IAV [9]), facilitating or hindering infection with a second pathogen. Moreover, a pathogen can change the **physiological environment** to potentiate a secondary infection by another pathogen. For instance, the replication of IAV in the respiratory epithelium reduces mucociliary clearance and damages epithelial cells, resulting in enhanced attachment and invasion of the pneumococcus [21]. **Changes in the respiratory tract microbiota** by an infection can lead to the acquisition of a new pathogen, or to overgrowth and invasion of an already present pathogen [27–29]. Lastly, changes in **host behaviors** caused by infection with a first pathogen can affect the risk of subsequent infection with another pathogen, even in the absence of within-host interaction between the two. Examples include self-isolation to reduce spread of disease in humans and reduced social contacts in infected animals [30,31].

Population-level impact

The biological mechanisms outlined above may affect population-level dynamics through their effects on different epidemiological parameters: susceptibility to infection, transmission of infection (characterized by the transmissibility and the duration of infection), and disease severity. Of note, multiple biological mechanisms can affect the same epidemiological parameter; conversely, the same biological mechanism can affect multiple epidemiological parameters. For example, IAV-induced epithelial damage and dampened pneumococcal clearance increase host susceptibility to the pneumococcus and disease severity in co-infection, as suggested by historical pandemics [32], demonstrated in experimental studies [19], and inferred from mechanistic modeling of epidemiological time-series [33,34]. The effect of interaction on transmission is more difficult to measure, as it is determined not only by the susceptibility of the exposed and the transmissibility of the infected, but also by the contact patterns between the two [35]. However, this effect can be approximated with animal models [36–38], or estimated with mathematical modeling based on epidemiological data [35]. Of note, as shown by the decline in various respiratory infections following the non-pharmaceutical interventions (NPIs) in the COVID-19 pandemic [35,39–42], transmission can be changed substantially by host behaviors.

## 3. Review of evidence on SARS-CoV-2 interactions

### 3.1 Experimental evidence from animal models

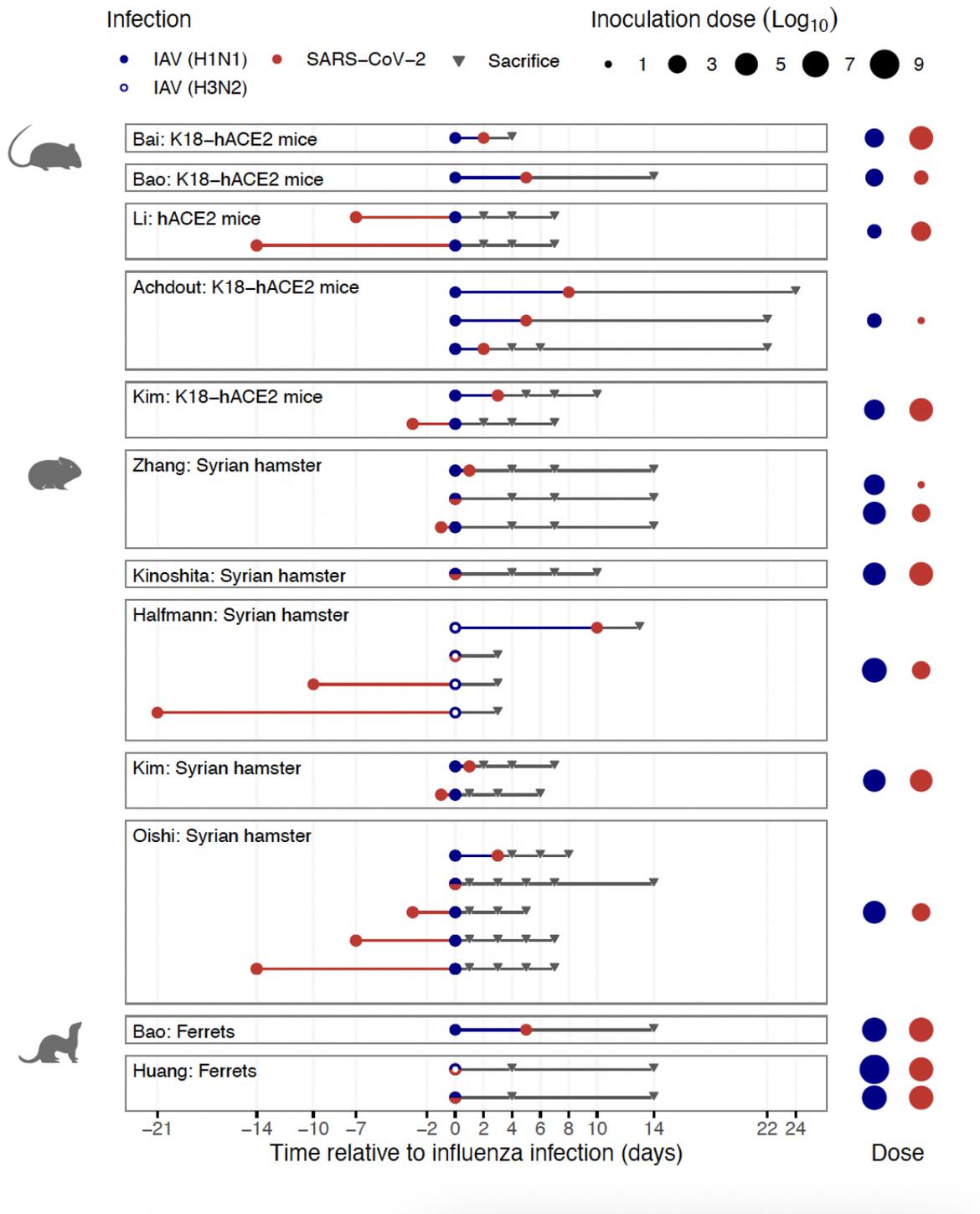

**Figure 2. Experimental designs of animal studies assessing the interaction between SARS-CoV-2 and influenza A virus[43–53].** The data from every study were extracted from the text,

the tables, or the figures; all the corresponding values were checked and are available in Tables S1 and S2.

Having proposed a framework to study interactions, we now review experimental studies on co-infections with SARS-CoV-2 in animal models. As of August 22th, 2022, we identified 14 publications [43–54,56–58]. We first review the 11 studies that focused on SARS-Cov_2 and non-attenuated IAV.

Experimental studies of co-infection with SARS-CoV-2 and non-attenuated IAV

As shown in Fig. 2, three different animal models were used (ferrets, hamsters, and mice) and the experimental designs varied substantially across the eleven studies, particularly in the sequence of infection, the time between infections (range: 0–21 days), and the follow-up duration (range: 3–24 days since first infection, 2–20 days since the second infection). Nine studies [43–47,49–52] examined co-infections with IAV preceding SARS-CoV-2, six [46,48–52] with SARS-CoV-2 preceding IAV, and five with simultaneous infections [46,47,49,51,53]. Of note, only three studies [46,49,51] compared all three infection sequences, and only four studies [45,48,49,51] compared different times between infections. Furthermore, the studies also varied widely in the inoculation dose (IAV range: $8\times10^1$–$1.3\times10^9$ PFU; SARS-CoV-2 range: $1\times10^1$–$7\times10^5$ PFU), with a single study [46] evaluating the effect of different doses. The studies used different IAV subtypes (H1N1 [43–48,50–53] and H3N2 [49,53]) and SARS-CoV-2 lineages (A [48,50,51,53], B [43,44,46,52], B.1 [45,49] and B1.1 [47]), as well as different strains within subtypes and lineages. Finally, only one study compared the effects of IAV (H1N1) and IAV (H3N2) [53]. Due to the limited number of studies and the large heterogeneity across them, we compare the results for SARS-CoV-2 and IAV (H1N1) co-infection only qualitatively.

As shown in Fig. 3A, the severity of mono-infection with either IAV or SARS-CoV-2 differed between animal models. In ferrets, mono-infection with IAV, but not with SARS-CoV-2, resulted in weight loss, while the opposite was observed in hamsters. In mice, however, both mono-infections generally caused weight loss. Also unlike the hamster and ferret models, mice can develop severe COVID-19 and die, so that this model was used in all studies that analyzed survival (Fig. 3B). On the whole, these results agree with earlier evidence of the advantages and limitations of different animal models for in vivo research on IAV and SARS-CoV-2 [59,60].

In all but one study, the effect of co-infection on disease severity was quantified by tracking changes in the animals' body weight. In mice and, to a lesser extent, in hamsters, animals co-infected suffered a higher maximal weight loss than animals mono-infected with either IAV or SARS-CoV-2 (Fig. 3A, Table S1). In ferrets, however, the maximum weight loss after co-infection was relatively comparable to that after IAV mono-infection. In keeping with the results based on weight loss, the three studies that measured survival (all using the mice model) found that co-infected animals either suffered higher mortality [45,50] or died faster [44] than mono-infected animals (Fig 3B, Table S1).

In contrast to the relatively consistent results on disease severity, the effect of co-infection on viral load—quantified as the ratio of viral load during co-infection to that during mono-infection— was more heterogeneous across studies (Fig 4, Table S2). In addition to the sources of heterogeneity outlined above, the studies varied in the technique used to quantify viral load (either RT-qPCR, plaque-based or median tissue culture infectious dose (TCID50) assays) and in the sample type (swabs or tissue) and location (lower respiratory tract (LRT) or upper respiratory tract (URT)). These differences may affect the inferred sign and strength of interaction: for example, the load of infectious viruses—which only plaque-based or TCID50 assays can quantify—in the URT is likely a more relevant proxy of transmissibility [61], but was measured in only six studies [45,46,49,51–53]. Overall, the effect size spanned six orders of magnitude and depended on the location of the body compartment sampled. In the LRT, the viral load of SARS-CoV-2 was generally reduced by preceding or simultaneous infection with IAV, but increased by subsequent infection with IAV in hamsters (Fig. 4A, left panel). The effect was more variable in mice, and inconclusive in ferrets because of a low number of studies. On the other hand, there was no obvious pattern in the viral load of IAV, regardless of infection order (Fig. 4A, right panel). In the upper respiratory tract, fewer studies assessed the effect of co-infection on viral load and their results were inconsistent (Fig. 4B).

Of note, several studies suggested time dependencies in co-infection outcomes. First, the maximum weight loss was typically observed 7–12 days post-infection ([44,45], Table S1), so that studies with shorter follow-up could under-estimate disease severity. Second, shorter time between infections was found to increase disease severity in two studies [45,48], (Fig. 3B, Table S1) and the effect on viral load in one study ([51], Fig. 4A, Table S2).

In conclusion, despite large heterogeneity and inconsistencies across the studies reviewed, the collective evidence from animal models shows that co-infection with IAV and SARS-CoV-2 causes more severe disease than mono-infection with either virus. Despite having clinical relevance, these results do not necessarily demonstrate a positive interaction. This is because the endpoints in all studies were non-specific, making it difficult to hypothesize the expected disease severity resulting from mere co-occurrence of two independent infections that do not interact. Virus-specific endpoints are therefore needed to conclusively demonstrate an interaction affecting disease severity. Despite the availability of such endpoints to assess the effect of co-infection on viral load, the collective evidence was inconclusive. A generally robust finding was that preceding or simultaneous infection with IAV reduced the viral load of SARS-CoV-2 in the LRT. However, only a few studies measured the viral load in the URT, which is likely a more relevant proxy of transmissibility [61]. Therefore, further studies will be needed to demonstrate the existence of interactions affecting susceptibility to, or transmissibility of, infection. In the design of such studies, we argue that the strength of evidence could be increased by varying the infectious dose and the infection order, and by considering different animal models.

Experimental studies of co-infection with SARS-CoV-2 and other pathogens

In addition to the previous studies, we identified three more experimental studies on SC2 co-infection [54,56,57]. One study found that administering live attenuated influenza A vaccine three days before SC2 infection reduced SC2 viral load in ferrets [54]. The second study observed that SC2 infection after, but not before, pneumococcal infection, increased the viral and bacterial loads, worsening disease severity and survival [57]. In contrast, the third study found that chronic infection with Mycobacterium tuberculosis inhibited SC2 viral load, decreasing disease severity [56].

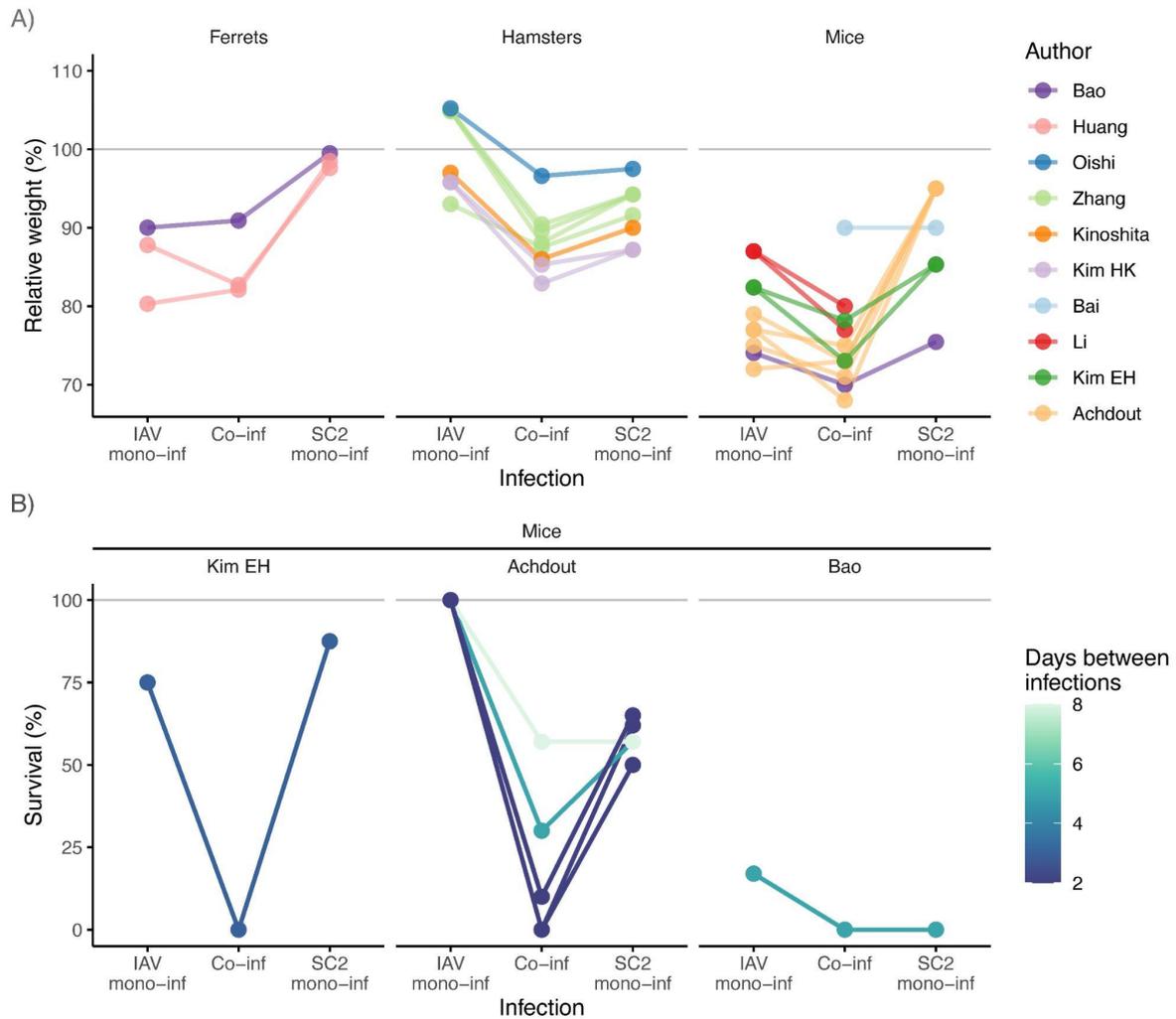

**Figure 3. Summary results from animal studies assessing the effect of co-infection with SARS-CoV-2 and influenza A virus (IAV) on disease severity [43–48,50–53].** In panel A, the y-axis values represent the weight relative to baseline, calculated when the maximal weight loss was observed (or, if the animals did not lose weight, when the maximum weight gain was observed). In panel B, the y-axis values represent the fraction of animals alive at the end of the experiment. The data from every study were extracted from the text, the tables, or the figures; all the corresponding values were checked and are available in Tables S1 and S2.

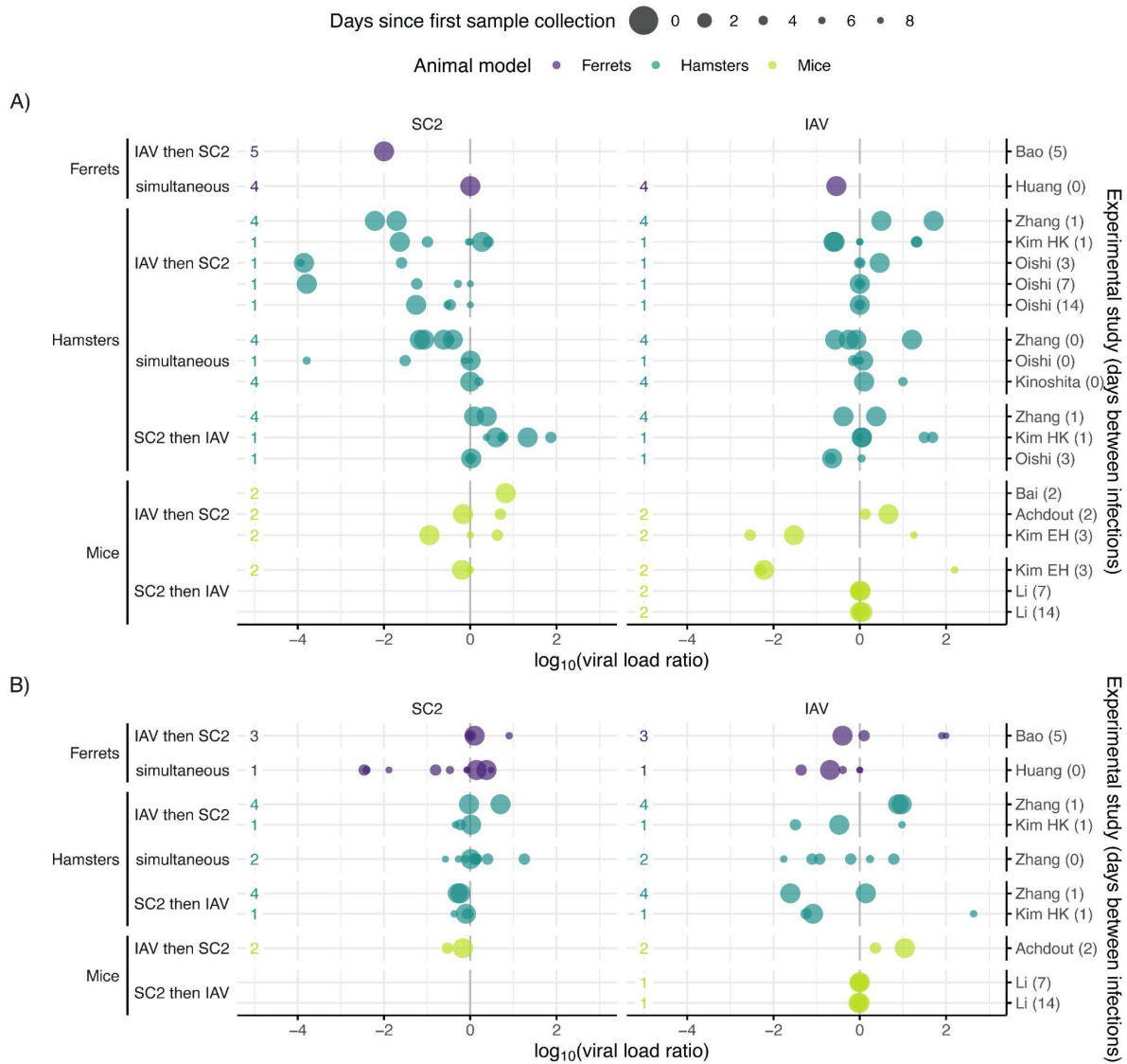

**Figure 4. Summary results from animal studies assessing the effect of co-infection with SARS-CoV-2 and influenza A virus (IAV) on viral loads [43–48,50–53].** The x-axis values represent the ratio of the viral load of SC2 (left panels) or IAV (right panels) during co-infection to that during mono-infection, in either the lower respiratory tract (panel A) or the upper respiratory tract (panel B). The colored numbers at the start of the x-axis represent the number of days after the last infection when the first sample was collected (for example, in the study by Huang the first sample was taken 4 days after simultaneous infection [0 days between infections]). The data from every study were extracted from the text, the tables, or the figures; all the corresponding values were checked and are available in Tables S1 and S2.

**3.2 Epidemiological evidence**

Although experimental studies using animal models can inform some of the components required to characterize pathogen interactions (Fig. 1), they are insufficient in predicting the public health impact of interaction in humans, for at least two reasons. First, animal models cannot fully recapitulate the biology of infection in humans, as illustrated by the ongoing search for an appropriate animal model representative of severe COVID-19 disease in humans [59]. Second, animal experimental studies may be under-powered to estimate relative risk of infection or severe disease in co- vs. mono-infected individuals. Hence, epidemiological studies remain indispensable to assess the significance of interaction in human populations. We reviewed the literature on SARS-CoV-2 and co-infections in human populations. The identified studies are classified into three categories: (1) studies that were based on co-infection prevalence, (2) studies that examined the association between non-COVID vaccines and COVID-19, and (3) studies that examined the association between prior respiratory infections and COVID-19.

Studies based on the detection of SARS-CoV-2 co-infections

Studies based on the detection of SARS-CoV-2 co-infections attempted to answer two research questions: (1) whether co-infection with other pathogens change the severity of COVID-19, or (2) whether the detection of other pathogens was associated with a change in SARS-CoV-2 detection.

Four meta-analyses addressed the first question. The first meta-analysis included only four studies, with large heterogeneity [62]. The second meta-analysis estimated a reduced mortality in patients co-infected with influenza from studies in China, (OR=0.51, 95% CI: 0.39–0.68, I2= 26.5%), but an increased mortality from studies outside China (OR=1.56, 95% CI: 1.12–2.19, I2= 1%) [63]. The two other meta-analyses reported higher mortality in SARS-CoV-2 co-infections compared with SARS-CoV-2 mono-infections. However, one of them did not provide information about the infection order [64]; the other provided separate estimates for when other respiratory pathogens were detected at the time of SARS-CoV-2 detection (OR=2.84, 95% CI: 1.42–5.66) or after (OR=3.54, 95% CI: 1.46–8.5), but pooled estimates for different age groups, healthcare settings (ICU and non-ICU), and pathogens (bacterial, viral and fungal) [65]. In general, all these studies require cautious interpretation, because confounders (such as comorbidities) may bias estimation.

Two studies used a test-negative design to address the second question, by comparing the prevalence of SARS-CoV-2 infection in groups infected vs. uninfected with another pathogen [77,78]. However, this approach can be inappropriate for two reasons. First, the prevalence of co-infections were likely under-estimated due to the prescription of empirical antibiotic treatment prior to microbiological investigation [68,74] and to diagnostic strategies favoring SARS-CoV-2 diagnosis [75]. Moreover, when simultaneous testing of multiple pathogens is limited, epidemics of co-circulating pathogens may artificially decrease the positivity fraction of SARS-CoV-2 [76]. Second, a less appreciated, but more essential problem of test-negative designs is that they can systematically underestimate the strength of interaction, and frequently infer the wrong sign of interaction for seasonal and emerging respiratory viruses [81]. These issues caution against simple and seemingly intuitive measures of pathogen interactions based on co-infection prevalence data, echoing earlier studies in infectious disease ecology and epidemiology [82–84].

Studies examining the association between non-COVID vaccination history and COVID-19

Since interacting pathogens form polymicrobial systems, interventions against any pathogen in such systems may theoretically affect the others. For example, if there is a positive interaction between a vaccine-preventable respiratory pathogen (e.g., IAV or the pneumococcus) and SARS-CoV-2, one may expect, with all else being equal, SARS-CoV-2-related outcomes to be higher in unvaccinated individuals. A systematic review [85] and two meta-analyses [86,87] have summarized a total of thirty articles on observational studies investigating the association of influenza vaccine and SARS-CoV-2 infections and outcomes. While the earlier systematic review (which included 12 studies) indicated that only some studies reported significantly inverse associations between influenza vaccination and SARS-CoV-2-related outcomes, the later meta-analyses (which included 16 and 23 studies respectively) found a significantly lower risk of SARS-CoV-2 infection associated with influenza vaccination (OR: 0.86, 95% CI: 0.81–0.91[86]; OR: 0.83, 95% CI: 0.76–0.90[87]).

In contrast to influenza vaccines, we found no systematic review that examined the association between pneumococcal conjugate vaccines (PCV) or pneumococcal polysaccharides vaccines (PPSV) and SARS-CoV-2 outcomes. Based on a literature review, we identified four studies—2 on PCV and PPSV [88–90], 1 on PCV only [91], and 1 on PPSV only [90] (Table S3). All three studies involving PPSV did not find conclusive evidence for association between PPSV history and SARS-CoV-2 related outcomes [88,89]. PCV was associated with protection against COVID-19 infection, hospitalization, and mortality among older adults in one cohort study [89], and against symptoms among SARS-CoV-2-infected children in another cohort study [91]. Although inconclusive, the association estimated in a case-control study [88] was consistent with that in the two cohort studies.

Findings from vaccine impact studies must be interpreted with caution when attempting to infer pathogen interactions. First, although numerous studies attempted to estimate the effect of various vaccines on COVID-19 outcomes, few accounted for healthy user bias, a common form of selection bias whereby more active health-seeking behaviors can be a source of confounding [92]. As acknowledged by [93] and [94], this is often a limitation in observational studies, as influenza vaccination is voluntary [94–97]. Second, even when epidemiological studies adopting more robust study designs (e.g., prospective cohort) and inference methods (e.g., Cox model with inverse propensity weighting) show that non-SARS-CoV-2 vaccines confer protection against SARS-CoV-2 [89], one cannot distinguish if such protection stems from hindering the positive interaction between two pathogens, or from the direct effect of the vaccine on SARS-CoV-2—for example via nonspecific immune responses such as trained innate immunity [98].

Studies examining the association between prior respiratory infections and COVID-19

Four observational studies reported the association of prior respiratory infections and COVID-19-related outcomes [99–102] (Table S4). Prior influenza infection was reported to be associated with increased COVID-19 susceptibility (OR: 3.07, 95% CI: 1.61–5.85 for 1–14 days prior, OR 1.91, 95% CI: 1.54–2.37 for 1–90 days prior) and severity (OR: 3.64, 95% CI: 1.55–9.21 for 1–14 days prior, OR: 3.59, 95% CI: 1.42–9.05 for 1–30 days prior) in a case-control study [99]. This evidence, suggestive of a positive interaction between influenza and SARS-CoV-2, is consistent with the findings from a mathematical modeling study [103]. Although a retrospective cohort study reported that prior infection with endemic human coronaviruses (hCoVs) was associated with protection against COVID-related ICU admission (OR: 0.1, 95% CI: 0.1–0.9) [100], a case-control study on serum samples from hospitalized COVID-19 patients found that hCoVs antibodies were not associated with protection against SARS-CoV-2 infections nor hospitalizations [101]. Regarding the impact of upper respiratory infections (URI), a retrospective cohort study found lower risk (OR: 0.76, 95% CI: 0.75, 0.77) of testing positive for SARS-CoV-2 among individuals URI diagnosed in the preceding year [102], while a case-control study found higher risk among individuals diagnosed with URI in the preceding 1–14 days (OR: 6.95, 95% CI: 6.38–7.58) and 1–90 days (OR: 2.70, 95% CI: 2.55–2.86) [99]. This discrepancy may be explained by the different URI definitions and time frames for exposure measurement, in addition to different study designs and included confounders. Because these studies provided information about the infection timeline, they offered stronger evidence to infer pathogen interactions than studies based on co-infection prevalence, and also more direct evidence than studies examining the association between non-COVID vaccines and COVID-19. Nevertheless, one should beware of how misclassification of exposure and imperfect control for confounding can limit such study designs in inferring pathogen interactions.

In summary, the evidence available from human population health data indicates that co-infection prevalence is largely variable, that influenza vaccines and PCVs may be associated with reduced risk of SARS-CoV-2, and that earlier influenza infection may be associated with higher risk of SARS-CoV-2 infection and disease severity. However, our review also highlighted the limitations in the current epidemiological literature, as many studies were prone to multiple biases, including confounding, and only very few [99–103] were designed to infer interaction.

4. **The need for transmission models to study epidemiological interactions**

As reviewed above, the results of epidemiological studies can be difficult to interpret and their designs insufficient to characterize all the components of interactions (Fig. 1). Arguably, more integrated approaches are therefore needed to capture the complexities described above and to determine how individual-level mechanisms of interaction translate into population-level dynamics of infection or disease.

Mathematical models of transmission offer a powerful and economical tool to study infectious disease dynamics [104]. To study pathogen interactions, such models can be formulated to incorporate biologically explicit mechanisms of interaction (in addition to the other elements of the framework proposed above) and predict their potentially non-linear effects on transmission dynamics [105]. By design, these models translate between scales, such that the population-level impact of a given individual-level mechanism of interaction can be simulated and predicted. To illustrate the relevance of such models, we formulated two basic models of SARS-CoV-2 interaction (see more details and equations in the Supplement), with either an endemic colonizing bacterium (e.g., the pneumococcus) or a respiratory virus causing seasonal epidemics (e.g., influenza). In both cases, we assumed a non-symmetric (i.e., no effect of SARS-CoV-2 on the other pathogen) interaction that caused a 1–5 fold (strength) decrease or increase (sign) of SARS-CoV-2 transmission (mechanism) from co-infected individuals (duration of interaction equal to the infectious period of the other pathogen). Importantly, the within-host processes causing interaction were not explicitly modeled, but their effects were represented by these interaction parameters. As shown in Fig. 5A, we find that even a moderately strong interaction with a commensal bacterium can substantially affect the dynamics of SARS-CoV-2, increasing its peak incidence by 3.5 fold for positive interaction when the prevalence of bacterial colonization reaches 50% of the population (as frequently observed in young children for the pneumococcus[106,107]). By contrast, an equal interaction with an epidemic virus is predicted to have a much smaller impact on the dynamics of SARS-CoV-2 (Fig. 5B). Of note, the maximal impact is predicted at intermediate levels of transmissibility of the epidemic virus, corresponding to maximal epidemic overlap with SARS-CoV-2 (Fig. 5B). This finding emphasizes a major difference between endemic and epidemic pathogens: for the latter, the impact of even strong interactions may remain subtle and manifest itself only after a prolonged period of co-circulation with SARS-CoV-2. Overall, these numerical experiments demonstrate the value of mathematical models to study interactions in a biologically explicit and comprehensive way and to predict their complex (and potentially unexpected) effects at the population level.

Although voluntarily over-simplified and used here only for illustrative and exploratory purposes, these models can be readily extended to add components relevant to SARS-CoV-2 epidemiology, such as age, vaccination, or temporal variations in transmission caused by new variants, seasonality, or changing control measures. In real-world applications, however, model parametrization can be a substantial challenge, as the values of many parameters may be neither directly observable nor fixed from empirical evidence. This problem is particularly salient for parameters characterizing interaction, whose values can be only partially inferred from experimental and epidemiological studies. To overcome this uncertainty, novel statistical inference techniques can be used to systematically compare the likelihood of different hypotheses about the mechanism, strength, and duration of interaction [108,109]. The potential of this approach is demonstrated by earlier successful applications [110,111], in particular to the system influenza–pneumococcus [33,34,112]. So far, however, few modeling studies have attempted to estimate the interactions of SARS-CoV-2 [103,113], presumably because of the near disappearance of many common diseases—caused, for example, by influenza and the pneumococcus [39,40]—after implementation of stringent control measures against COVID-19. In light of the likely relaxation of these measures and the ensuing increase in co-circulating pathogens, we anticipate that confronting mathematical models with detailed epidemiological surveillance data will increasingly provide valuable insights into the interactions of SARS-CoV-2.

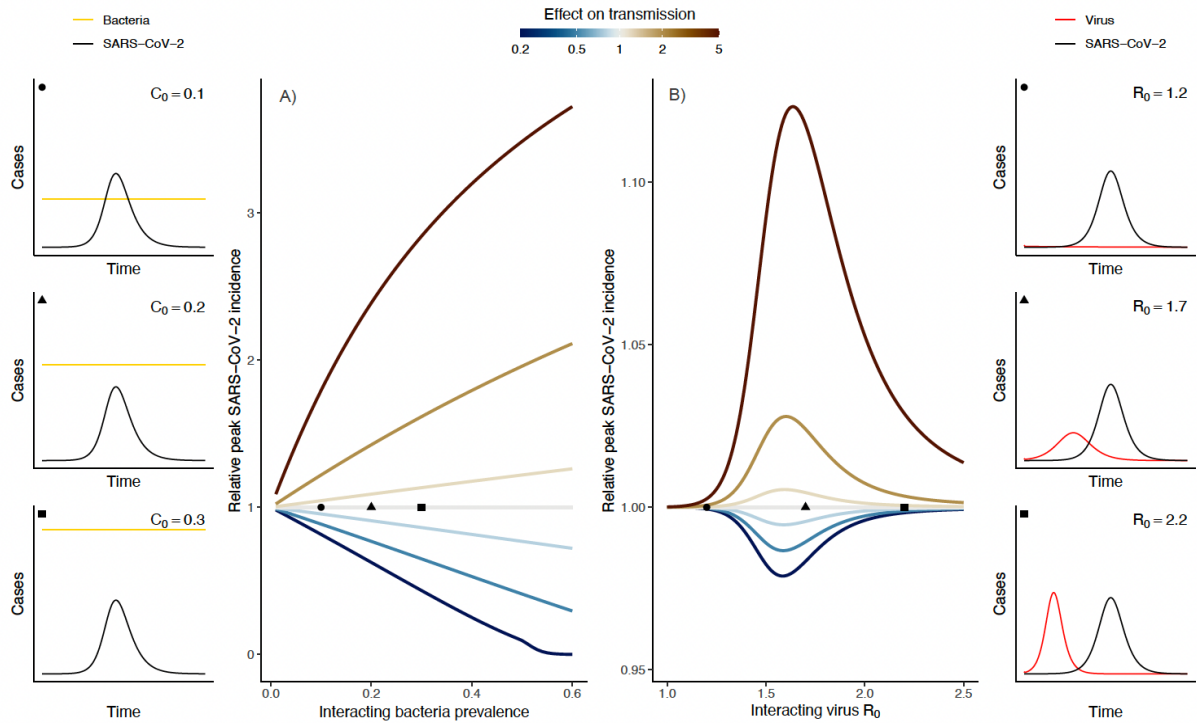

**Figure 5. Potential impact of interaction on SARS-CoV-2 dynamics from mathematical models of SARS-CoV-2 interaction.** A: Interaction with colonizing bacteria (e.g., *Streptococcus pneumoniae*). B: Interaction with a seasonal virus (e.g., influenza A virus). Insets represent three example simulations for each of the two models, varying the prevalence of bacterial colonization ($C_0$) and the basic reproduction number ($R_0$) of the interacting virus. Note, the vertical axes are on different scales, showing the more pronounced impact of interactions with endemic colonizing bacteria. The data presented are primary data, generated from illustrative models designed for the purpose of this review; full model details are included in appendix 5.

## 5. Conclusion

As population immunity against COVID-19 accrues in many regions worldwide, it is critical to understand the factors that will affect the future transmission dynamics of SARS-CoV-2 [2]. Here, we proposed that interactions with co-circulating pathogens will be such a key factor. Indeed, such interactions may have notable public health implications, in particular for forecasting and controlling SARS-CoV-2 epidemics and for predicting the indirect impact of vaccines. The scientific implications of interaction are also notable and may lead to considering SARS-CoV-2 as part of polymicrobial systems whose individual components cannot be well studied separately.

Despite the relevance of interaction, our review identified only a few experimental studies in animal models, with markedly different designs and the majority focusing on SARS-CoV-2 interaction with IAV. A robust finding from our comparative analysis is that SARS-CoV-2 and IAV co-infections can increase the severity of COVID-19. By contrast, the estimated effect of co-infection on influenza and SARS-CoV-2 viral loads differed markedly across studies, presumably because of the heterogeneous designs and methods to quantify viral load. Perhaps less surprisingly, the design and the results of epidemiological studies on interaction also varied widely. Although previous influenza vaccination was generally associated with a reduced risk of COVID-19, this finding alone does not necessarily provide evidence of positive interaction and may be equally well explained by direct, non-specific effects of influenza vaccines on host immunity. Nevertheless, the evidence from epidemiological [99] and mathematical modeling [103] studies suggest a facilitatory effect of previous influenza infection on SARS-CoV-2 infection. Besides influenza, few studies investigated the impact of other pathogens, in particular other major respiratory viruses like RSV and rhinoviruses, or colonizing bacteria like the pneumococcus [114]. In particular, research specific to interactions with endemic bacteria are called for, because—as illustrated by our simple model—these could substantially affect the dynamics of SARS-CoV-2. As a more general conclusion, our review suggests the urgent need for further experimental and epidemiological studies to unequivocally infer SARS-CoV-2 interactions.

Altogether, our review highlights the significant gaps that remain in our knowledge of SARS-CoV-2 interactions. The general framework proposed to dissect interaction may therefore be useful to guide further research in this field. We argue that mathematical models of transmission offer an intrinsically efficient way to incorporate this framework. Hence, we submit that such models—designed with a multi-disciplinary perspective that integrates evidence across scientific fields—will prove to be valuable tools to decipher the interactions of SARS-CoV-2.

**Declarations**

**Data availability statement:** The R codes to implement the models and reproduce Fig. 5 are available at: https://github.com/egoult/pathogen_coinfections

**Competing interests:** MDdC received postdoctoral funding (2017–2019) from Pfizer and consulting fees from GSK. LO reports research grants from Pfizer and Sanofi-Pasteur through her institution. All other authors declare no competing interests.

**Authors Contributions:** AW, LABG, EG, and MDdC wrote the first draft of the manuscript. LABG and MB conducted the review of experimental studies. AW conducted the review of epidemiological studies. EG and AK designed and implemented the transmission models of SARS-CoV-2 interactions. SCK and LO conducted bibliographical research and provided content expertise. MDdC conceived of the study design and oversaw the analysis. All authors helped draft, and approved the final version of, the manuscript.

**Acknowledgements** We thank Molly Sauter for help during the initial rounds of review. We also thank Dr. Hagit Achdout and Dr. Tomer Israely (Department of Infectious Diseases, Israel Institute for Biological Research, Israel), and Dr. Longding Liu (Institute of Medical Biology Chinese Academy of Medical Sciences, China) for their helpful discussion.

**List of supplementary materials**

Appendix 1: Table S1. An overview of the experimental designs and results on disease severity, measured as maximal body mass loss or survival at experiment end, from the reviewed studies assessing the interaction between SARS-CoV-2 and influenza A virus (IAV).

Appendix 2: Table S2. An overview of the experimental designs and results on viral load, measured in the upper or lower respiratory tract, from the reviewed studies assessing the interaction between SARS-CoV-2 and influenza A virus (IAV).

Appendix 3: Table S3. Observational studies examining the association between pneumococcal vaccination history and COVID-19.

Appendix 4: Table S4. Observational studies examining the association between prior respiratory infections and COVID-19.

Appendix 5: Model details (including Figure S1: Schematic of the bacteria-virus interaction model; Table S5. Parameters used for S. pneumoniae - SARS-CoV-2 interaction model; Figure S2: Schematic of the virus-virus interactions model; and Table S6. Parameters used for influenza A - SARS-CoV-2 interaction model).

# The interactions of SARS-CoV-2 with co-circulating pathogens: Epidemiological implications and current knowledge gaps

# Supplementary Materials

Anabelle Wong[1,2,*], Laura Andrea Barrero Guevara[1,2,*], Elizabeth Goult[1,*], Michael Briga[1], Sarah C. Kramer[1], Aleksandra Kovacevic[3,4], Lulla Opatowski[3,4], Matthieu Domenech de Cellès[1,**]

1. Infectious Disease Epidemiology group, Max Planck Institute for Infection Biology, Charitéplatz 1, 10117 Berlin, Germany.
2. Institute of Public Health, Charité–Universitätsmedizin Berlin, Charitéplatz 1, 10117 Berlin, Germany.
3. Epidemiology and Modelling of Antibiotic Evasion, Institut Pasteur, Paris, France
4. Anti-infective Evasion and Pharmacoepidemiology Team, CESP, Université Paris-Saclay, Université de Versailles Saint-Quentin-en-Yvelines, INSERM U1018 Montigny-le-Bretonneux, France.

\* These authors contributed equally.

\*\* Corresponding author: Dr. Matthieu Domenech de Cellès, Max Planck Institute for Infection Biology, Charitéplatz 1, Campus Charité Mitte, 10117 Berlin, Germany

**Table of Contents**

**Appendix 1: Table S1. An overview of the experimental designs and results on disease severity, measured as maximal body mass loss or survival at experiment end, from the reviewed studies assessing the interaction between SARS-CoV-2 and influenza A virus (IAV).** Values were taken from tables or text, or when these were not available, extracted from the figures using the program PlotDigitizer [12].

| Author | Year | Animal model | Sex | IAV (strain) | SC2 (Pangolin lineage) | SC2 infection order | Days btwn infections | Sample size per group (IAV-SC2-Coinf) | Inoculation dose IAV - SC2 extracted | Inoculation dose IAV - SC2 [PFU][13] | Follow-up days | Max body mass loss IAV | Day IAV | SC2 | Day SC2 | Coinf | Day Coinf | Figure/Table | Survival at experiment end IAV | SC2 | Coinf | Figure/Table |
|---|---|---|---|---|---|---|---|---|---|---|---|---|---|---|---|---|---|---|---|---|---|---|
| Bai et al. [1] | 2021 | K18-hACE2 mice | male | H1N1 (A/Sichuan/01/2009) | SC2 (B) | 2nd | 2 | 4-4-4 | 2X10^3 PFU - 3X10^5 PFU | 2X10^3 - 3X10^5 | 4 | NA | NA | -10,0% | 4 | -10,0% | 4 | Fig 2B | NA | NA | NA | NA |
| Bao et al. [2] | 2021 | Ferrets | male | H1N1 (A/California/07/2009) | SC2 (B) | 2nd | 5 | 6-6-6 | 1X10^6 CCID50 - 1X10^6 CCID50 | 7X10^5 - 7X10^5 | 10 | -10,0% | 6 | -0,5% | 3 | -9,1% | 6 | Fig1B | NA | NA | NA | NA |
| Bao et al. [2] | 2021 | K18-hACE2 mice | female | H1N1 (A/California/07/2009) | SC2 (B) | 2nd | 5 | 6-6-6 | 1X10^3 CCID50 - 1X10^2 CCID50 | 7X10^2 - 7X10^1 | 14 | -26,0% | 9 | -24,5% | 8 | -30,0% | 6 | Fig 4C | 17,0% | 0,0% | 0,0% | Fig 4B/C |
| Zhang et al. [3] | 2021 | Syrian hamster | male + female | H1N1 (A/Hong Kong/415742/2009) | SC2 (B) | simultaneous | 0 | 3-3-3 | 1X10^5 PFU - 1X10^3 PFU | 1X10^5 - 1X10^3 | 4 | -7,0% | 2 | -8,4% | 4 | -12,5% | 4 | Fig 2A | NA | NA | NA | NA |
| Zhang et al. [3] | 2021 | Syrian hamster | male + female | H1N1 (A/Hong Kong/415742/2009) | SC2 (B) | simultaneous | 0 | 3-3-3 | 1X10^4 PFU - 1X10^1 PFU | 1X10^4 - 1X10^1 | 4 | 4,9% | 4 | -5,8% | 4 | -9,6% | 4 | Fig 3A, Table 1 | NA | NA | NA | NA |
| Zhang et al. [3] | 2021 | Syrian hamster | male + female | H1N1 (A/Hong Kong/415742/2009) | SC2 (B) | 1st | -1 | 3-3-3 | 1X10^4 PFU - 1X10^1 PFU | 1X10^4 - 1X10^1 | 4 | 4,9% | 4 | -5,8% | 4 | -11,9% | 4 | Fig 3A, Table 1 | NA | NA | NA | NA |
| Zhang et al. [3] | 2021 | Syrian hamster | male + female | H1N1 (A/Hong Kong/415742/2009) | SC2 (B) | 2nd | 1 | 3-3-3 | 1X10^4 PFU - 1X10^1 PFU | 1X10^4 - 1X10^1 | 4 | 4,9% | 4 | -5,8% | 4 | -10,4% | 4 | Fig 3A, Table 1 | NA | NA | NA | NA |
| Achdout et al. [4] | 2021 | K18-hACE2 mice | female | H1N1 (A/Puerto Rico/8/1934, PR8) | SC2 (B.1) | 2nd | 2 | 10-13-11 | 8X10^1 PFU - 1X10^1 PFU | 8X10^1 - 1X10^1 | 20-24 | -25,0% | 10 | -5,0% | 9 | -29,0% | 8 | Fig 1A | 100,0% | 62,0% | 0,0% | Fig 1B |
| Achdout et al. [4] | 2021 | K18-hACE2 mice | female | H1N1 (A/Puerto Rico/8/1934, PR8) | SC2 (B.1) | 2nd | 5 | 11-14-7 | 8X10^1 PFU - 1X10^1 PFU | 8X10^1 - 1X10^1 | 20-24 | -21,0% | 8 | -5,0% | 11 | -27,0% | 10 | Fig 1C | 100,0% | 57,0% | 30,0% | Fig 1D |
| Achdout et al. [4] | 2021 | K18-hACE2 mice | female | H1N1 (A/Puerto Rico/8/1934, PR8) | SC2 (B.1) | 2nd | 8 | 8-14-7 | 8X10^1 PFU - 1X10^1 PFU | 8X10^1 - 1X10^1 | 20-24 | -23,0% | 9 | -5,0% | 14 | -25,0% | 8 | Fig 1E | 100,0% | 57,0% | 57,0% | Fig 1F |
| Achdout et al. [4] | 2021 | K18-hACE2 mice | female | H1N1 (A/Puerto Rico/8/1934, PR8) | SC2 (B.1) | 2nd | 2 | 4-6-4 | 8X10^1 PFU - 1X10^1 PFU | 8X10^1 - 1X10^1 | 20-24 | -23,0% | 9 | -5,0% | 9 | -32,0% | 8 | Fig 4C | 100,0% | 50,0% | 0,0% | Fig 4B |
| Achdout et al. [4] | 2021 | K18-hACE2 mice | female | H1N1 (A/Puerto Rico/8/1934, PR8) | SC2 (B.1) | 2nd | 2 | 7-9-9 | 8X10^1 PFU - 1X10^1 PFU | 8X10^1 - 1X10^1 | 20-24 | -28,0% | 10 | -5,0% | 7 | -27%* | 7 | Fig 4F | 100,0% | 65,0% | 10,0% | Fig 4E |
| Kinoshita et al. [5] | 2021 | Syrian hamster | female | H1N1 (A/Puerto Rico/8/1934, PR8) | SC2 (B.1.1) | simultaneous | 0 | 6-6-6 | 1X10^5 PFU - 3X10^5 PFU | 1X10^5 - 3X10^5 | 10 | -3,0% | 3 | -10,0% | 6 | -14,0% | 7 | Fig 1A | NA | NA | NA | NA |
| Li et al. [6] | 2021 | hACE2 mice | female | H1N1 (A/Puerto Rico/8/1934, PR8) | SC2 (A) | 1st | -7 | 9-9-9 | 1X10^2 CCID50 - 5X10^3 CCID50 | 7X10^1 - 3.5X10^3 | 7 | -13,0% | 7 | NA | NA | -23,0% | 7 | Fig 3A | NA | NA | NA | NA |
| Li et al. [6] | 2021 | hACE2 mice | female | H1N1 (A/Puerto Rico/8/1934, PR8) | SC2 (A) | 1st | -14 | 9-9-9 | 1X10^2 CCID50 - 5X10^3 CCID50 | 7X10^1 - 3.5X10^3 | 14 | -13,0% | 7 | NA | NA | -20,0% | 7 | Fig 3A | NA | NA | NA | NA |
| Halfmann et al. [7] | 2021 | Syrian hamster | female | H3N2 (A/Tokyo/UT-IMS3-1/2014) | SC2 (B.1) | simultaneous | 0 | NA | 1X10^6 PFU - 1X10^3 PFU | 1X10^6 - 1X10^3 | NA | NA | NA | NA | NA | NA | NA | NA | NA | NA | NA | NA |
| Halfmann et al. [7] | 2021 | Syrian hamster | female | H3N2 (A/Tokyo/UT-IMS3-1/2014) | SC2 (B.1) | 1st | -10 | NA | 1X10^6 PFU - 1X10^3 PFU | 1X10^6 - 1X10^3 | NA | NA | NA | NA | NA | NA | NA | NA | NA | NA | NA | NA |
| Halfmann et al. [7] | 2021 | Syrian hamster | female | H3N2 (A/Tokyo/UT-IMS3-1/2014) | SC2 (B.1) | 2nd | 10 | NA | 1X10^6 PFU - 1X10^3 PFU | 1X10^6 - 1X10^3 | NA | NA | NA | NA | NA | NA | NA | NA | NA | NA | NA | NA |
| Kim et al. [8] | 2022 | K18-hACE2 mice | female | H1N1 (A/California/04/2009) | SC2 (A) | 1st | -3 | 29-29-26 | 1X10^4 TCID50 - 1X10^5.5 TCID50 | 7X10^3 - 7X10^4.5 | 10 | -17,6% | 9 | -14,7% | 7 | -27,0% | 10 | Fig 1B | 75,0% | 87,5% | 0,0% | Fig 1C |
| Kim et al. [8] | 2022 | K18-hACE2 mice | female | H1N1 (A/California/04/2009) | SC2 (A) | 2nd | 3 | 29-29-26 | 1X10^4 TCID50 - 1X10^5.5 TCID50 | 7X10^3 - 7X10^4.5 | 10 | -17,6% | 9 | -14,7% | 7 | -21,8% | 10 | Fig 1B | 75,0% | 87,5% | 0,0% | Fig 1C |
| Huang et al. [9] | 2022 | Ferrets | female | H1N1 (A/California/07/2009) | SC2 (A) | simultaneous | 0 | 4 | 1X10^6 PFU - 5X10^5 PFU | 1X10^6 - 5X10^5 | 14 | -12,2% | 6 | -2,4% | 14 | -17,3% | 7 | Fig 1B | NA | NA | NA | NA |
| Huang et al. [9] | 2022 | Ferrets | female | H3N2 (A/Kansas/14/2017) | SC2 (A) | simultaneous | 0 | 4 | 1.3X10^9 PFU - 5X10^5 PFU | 1.3X10^9 - 5X10^5 | 14 | -3,1% | 5 | -2,4% | 14 | -5,6% | 7 | Fig 1B | NA | NA | NA | NA |
| Huang et al. [9] | 2022 | Ferrets | female | H1N1 (A/California/07/2009) | SC2 (A) | simultaneous | 0 | 4 | 1X10^6 PFU - 5X10^5 PFU | 1X10^6 - 5X10^5 | 13 | -19,7% | 5 | -1,5% | 5 | -17,9% | 7 | Fig 5B/C/F | NA | NA | NA | NA |
| Huang et al. [9] | 2022 | Ferrets | female | H3N2 (A/Kansas/14/2017) | SC2 (A) | simultaneous | 0 | 4 | 1.3X10^9 PFU - 5X10^5 PFU | 1.3X10^9 - 5X10^5 | 13 | -3,3% | 5 | -1,5% | 5 | -5,4% | 14 | Fig 5D/E/F | NA | NA | NA | NA |
| Kim et al. [10] | 2022 | Syrian hamster | male | H1N1 (A/California/04/2009) | SC2 (B) | 1st | -1 | Unreported | 1X10^5 TCID50 - 1X10^5 TCID50 | 7X10^4 - 7X10^4 | 7 | -4,2% | 1 | -12,8% | 5 | -14,7% | 3 | Fig 1B | NA | NA | NA | NA |
| Kim et al. [10] | 2022 | Syrian hamster | male | H1N1 (A/California/04/2009) | SC2 (B) | 2nd | 1 | Unreported | 1X10^5 TCID50 - 1X10^5 TCID50 | 7X10^4 - 7X10^4 | 7 | -4,2% | 1 | -12,8% | 5 | -17,1% | 7 | Fig 1B | NA | NA | NA | NA |
| Oishi et al. [11] | 2022 | Syrian hamster | male | H1N1 (A/California/04/2009) | SC2 (A) | simultaneous | 0 | 8-8-8 | 1X10^5 PFU - 1x10^3 PFU | 1X10^5 - 1x10^3 | 8 | 5,2% | 3 | -2,5% | 3 | -3,4% | 3 | Fig 2D | NA | NA | NA | NA |
| Oishi et al. [11] | 2022 | Syrian hamster | male | H1N1 (A/California/04/2009) | SC2 (A) | 1st | -3 | NA | 1X10^5 PFU - 1x10^3 PFU | 1X10^5 - 1x10^3 | NA | NA | NA | NA | NA | NA | NA | NA | NA | NA | NA | NA |
| Oishi et al. [11] | 2022 | Syrian hamster | male | H1N1 (A/California/04/2009) | SC2 (A) | 2nd | 3 | NA | 1X10^5 PFU - 1x10^3 PFU | 1X10^5 - 1x10^3 | NA | NA | NA | NA | NA | NA | NA | NA | NA | NA | NA | NA |
| Oishi et al. [11] | 2022 | Syrian hamster | male | H1N1 (A/California/04/2009) | SC2 (A) | 2nd | 7 | NA | 1X10^5 PFU - 1x10^3 PFU | 1X10^5 - 1x10^3 | NA | NA | NA | NA | NA | NA | NA | NA | NA | NA | NA | NA |
| Oishi et al. [11] | 2022 | Syrian hamster | male | H1N1 (A/California/04/2009) | SC2 (A) | 2nd | 14 | NA | 1X10^5 PFU - 1x10^3 PFU | 1X10^5 - 1x10^3 | NA | NA | NA | NA | NA | NA | NA | NA | NA | NA | NA | NA |

**Abbreviations** K18-hACE2 mice: transgenic mice expressing human angiotensin-converting enzyme 2 (hACE2) controlled by the human cytokeratin 18 promoter, PFU: plaque-forming unit, CCID50: Cell culture infectious dose 50%, TCID50: Tissue culture infectious dose 50%.

*****Remark** Viral dose concentrations in PFU following Daelemans et al. [10] protocol (1 CCID50 = 0.7 PFU).

Appendix 2: Table S2. An overview of the experimental designs and results on viral load, measured in the upper or lower respiratory tract, from the reviewed studies assessing the interaction between SARS-CoV-2 and influenza A virus (IAV). Values were obtained from tables or text, or when these were not available, from figures

| Author | Year | Animal model | Sex | IAV (strain) | SC2 (Pangolin lineage) | SC2 infection order | Days btwn infections | Sample size (per group) | Inoculation dose | Viral load in lower respiratory tract ||||||| Viral load in upper respiratory tract ||||||| Quantification Unit | Quantification Method |
|---|---|---|---|---|---|---|---|---|---|---|---|---|---|---|---|---|---|---|---|---|---|---|---|---|---|
| | | | | | | | | | | Tissue | Sample | Sampling day (dpi) | Coinf (IAV) | Monoinf (IAV) | Coinf (SC2) | Monoinf (SC2) | Figure/Table | Tissue | Sample | Sampling day (dpi) | Coinf (IAV) | Monoinf (IAV) | Coinf (SC2) | Monoinf (SC2) | Figure/Table | | |
| Bai et al. [1] | 2021 | K18-hACE2 mice | male | H1N1 (A/Sichuan/01/2009) | SC2 (B) | 2nd | 2 | 3 | See table S1 | Lung | Tissue | 2 | NA | NA | 6,60 | 1,00 | Fig 2D | NA | NA | NA | NA | NA | NA | NA | NA | gc/GAPDH | RT-qPCR |
| Bao et al. [2] | 2021 | Ferrets | male | H1N1 (A/California/07/2009) | SC2 (B) | 2nd | 5 | 4 | See table S1 | Lung | Tissue | 5 | NA | NA | 1,20 | 3,20 | Fig 2C | Throat | Swabs | 3 | 5,80 | 6,20 | 5,50 | 5,40 | Fig 2A/B | log10 gc/mL | RT-qPCR |
| Bao et al. [2] | 2021 | Ferrets | male | H1N1 (A/California/07/2009) | SC2 (B) | 2nd | 5 | 4 | See table S1 | NA | NA | NA | NA | NA | NA | NA | NA | Throat | Swabs | 5 | 5,40 | 5,30 | 4,90 | 4,00 | Fig 2A/B | log10 gc/mL | RT-qPCR |
| Bao et al. [2] | 2021 | Ferrets | male | H1N1 (A/California/07/2009) | SC2 (B) | 2nd | 5 | 4 | See table S1 | NA | NA | NA | NA | NA | NA | NA | NA | Throat | Swabs | 8 | 4,80 | 2,90 | 0,90 | 0,00 | Fig 2A/B | log10 gc/mL | RT-qPCR |
| Bao et al. [2] | 2021 | Ferrets | male | H1N1 (A/California/07/2009) | SC2 (B) | 2nd | 5 | 4 | See table S1 | NA | NA | NA | NA | NA | NA | NA | NA | Throat | Swabs | 10 | 2,00 | 0,00 | 0,00 | 0,00 | Fig 2A/B | log10 gc/mL | RT-qPCR |
| Zhang et al. [3] | 2021 | Syrian hamster | male+female | H1N1 (A/Hong Kong/415742/2009) | SC2 (B) | simultaneous | 0 | 3 | See table S1: High dose | Lung | Tissue | 4 | 4,78 | 4,86 | 5,21 | 5,82 | Fig 2C/D | Nasal Turbinates | Tissue | 4 | 4,79 | 4,00 | 5,80 | 4,55 | Fig 2C/D | log10 PFU/mL | Plaque-based |
| Zhang et al. [3] | 2021 | Syrian hamster | male+female | H1N1 (A/Hong Kong/415742/2009) | SC2 (B) | simultaneous | 0 | 3 | See table S1: High dose | Lung | Tissue | 4 | -1,37 | -1,12 | 0,04 | 0,45 | Fig 2C/D | Nasal Turbinates | Tissue | 4 | -0,94 | 0,16 | 1,56 | 1,15 | Fig 2C/D | log10 gc/beta-actin | RT-qPCR |
| Zhang et al. [3] | 2021 | Syrian hamster | male+female | H1N1 (A/Hong Kong/415742/2009) | SC2 (B) | simultaneous | 0 | 3 | See table S1: Low dose | Lung | Tissue | 4 | 3,82 | 2,61 | 3,55 | 4,62 | Table 1 | Nasal Turbinates | Tissue | 4 | 2,24 | 3,17 | 5,61 | 5,48 | Table 1 | log10 PFU/mL | Plaque-based |
| Zhang et al. [3] | 2021 | Syrian hamster | male+female | H1N1 (A/Hong Kong/415742/2009) | SC2 (B) | simultaneous | 0 | 3 | See table S1: Low dose | Lung | Tissue | 4 | 3,41 | 3,98 | 4,19 | 5,36 | Table 1 | Nasal Turbinates | Tissue | 4 | 3,25 | 3,46 | 5,14 | 5,00 | Table 1 | log10 PFU/mL | Plaque-based |
| Zhang et al. [3] | 2021 | Syrian hamster | male+female | H1N1 (A/Hong Kong/415742/2009) | SC2 (B) | 1st | -1 | 3 | See table S1: Low dose | Lung | Tissue | 4 | 2,23 | 2,61 | 4,71 | 4,62 | Table 1 | Nasal Turbinates | Tissue | 4 | 1,56 | 3,17 | 5,25 | 5,48 | Table 1 | log10 PFU/mL | Plaque-based |
| Zhang et al. [3] | 2021 | Syrian hamster | male+female | H1N1 (A/Hong Kong/415742/2009) | SC2 (B) | 1st | -1 | 3 | See table S1: Low dose | Lung | Tissue | 4 | 4,36 | 3,98 | 5,74 | 5,36 | Table 1 | Nasal Turbinates | Tissue | 4 | 3,60 | 3,46 | 4,71 | 5,00 | Table 1 | log10 PFU/mL | Plaque-based |
| Zhang et al. [3] | 2021 | Syrian hamster | male+female | H1N1 (A/Hong Kong/415742/2009) | SC2 (B) | 2nd | 1 | 3 | See table S1: Low dose | Lung | Tissue | 4 | 4,32 | 2,61 | 2,41 | 4,62 | Table 1 | Nasal Turbinates | Tissue | 4 | 4,06 | 3,17 | 5,45 | 5,48 | Table 1 | log10 PFU/mL | Plaque-based |
| Zhang et al. [3] | 2021 | Syrian hamster | male+female | H1N1 (A/Hong Kong/415742/2009) | SC2 (B) | 2nd | 1 | 3 | See table S1: Low dose | Lung | Tissue | 4 | 4,48 | 3,98 | 3,65 | 5,36 | Table 1 | Nasal Turbinates | Tissue | 4 | 4,43 | 3,46 | 5,70 | 5,00 | Table 1 | log10 PFU/mL | Plaque-based |
| Zhang et al. [3] | 2021 | Syrian hamster | male+female | H1N1 (A/Hong Kong/415742/2009) | SC2 (B) | simultaneous | 0 | 3 | See table S1: Low dose | Lung | Tissue | 7 | 0/NA | -3,74 | -2,74 | -2,27 | Fig 4G | Nasal Turbinates | Tissue | 7 | -4,75 | -2,99 | 0,14 | 0,41 | Fig 4G | log10 gc/beta-actin | RT-qPCR |
| Zhang et al. [3] | 2021 | Syrian hamster | male+female | H1N1 (A/Hong Kong/415742/2009) | SC2 (B) | simultaneous | 0 | 3 | See table S1: Low dose | Lung | Tissue | 14 | 0/NA | 0/NA | 0/NA | 0/NA | Fig 4G | Nasal Turbinates | Tissue | 14 | 0/NA | 0/NA | -2,55 | -2,67 | Fig 4G | log10 gc/beta-actin | RT-qPCR |
| Zhang et al. [3] | 2021 | Syrian hamster | male+female | H1N1 (A/Hong Kong/415742/2009) | SC2 (B) | simultaneous | 0 | 3 | See table S1: Low dose | NA | NA | NA | NA | NA | NA | NA | NA | Nasal Turbinates | Swabs | 2 | 3,53 | NA | 5,38 | 5,37 | Fig 4H | log10 gc/beta-actin | RT-qPCR |
| Zhang et al. [3] | 2021 | Syrian hamster | male+female | H1N1 (A/Hong Kong/415742/2009) | SC2 (B) | simultaneous | 0 | 3 | See table S1: Low dose | NA | NA | NA | NA | NA | NA | NA | NA | Nasal Turbinates | Swabs | 4 | 2,81 | NA | 5,38 | 5,29 | Fig 4H | log10 gc/beta-actin | RT-qPCR |
| Zhang et al. [3] | 2021 | Syrian hamster | male+female | H1N1 (A/Hong Kong/415742/2009) | SC2 (B) | simultaneous | 0 | 3 | See table S1: Low dose | NA | NA | NA | NA | NA | NA | NA | NA | Nasal Turbinates | Swabs | 6 | 3,26 | 3,02 | 3,79 | 3,90 | Fig 4H | log10 gc/beta-actin | RT-qPCR |
| Zhang et al. [3] | 2021 | Syrian hamster | male+female | H1N1 (A/Hong Kong/415742/2009) | SC2 (B) | simultaneous | 0 | 3 | See table S1: Low dose | NA | NA | NA | NA | NA | NA | NA | NA | Nasal Turbinates | Swabs | 8 | 0/NA | NA | 3,74 | 4,32 | Fig 4H | log10 gc/beta-actin | RT-qPCR |
| Zhang et al. [3] | 2021 | Syrian hamster | male+female | H1N1 (A/Hong Kong/415742/2009) | SC2 (B) | simultaneous | 0 | 3 | See table S1: Low dose | NA | NA | NA | NA | NA | NA | NA | NA | Nasal Turbinates | Swabs | 10 | 0/NA | 0/NA | 3,36 | 0/NA | Fig 4H | log10 gc/beta-actin | RT-qPCR |
| Zhang et al. [3] | 2021 | Syrian hamster | male+female | H1N1 (A/Hong Kong/415742/2009) | SC2 (B) | simultaneous | 0 | 3 | See table S1: Low dose | NA | NA | NA | NA | NA | NA | NA | NA | Nasal Turbinates | Swabs | 12 | 0/NA | 0/NA | 0/NA | 0/NA | Fig 4H | log10 gc/beta-actin | RT-qPCR |
| Zhang et al. [3] | 2021 | Syrian hamster | male+female | H1N1 (A/Hong Kong/415742/2009) | SC2 (B) | simultaneous | 0 | 3 | See table S1: Low dose | NA | NA | NA | NA | NA | NA | NA | NA | Nasal Turbinates | Swabs | 14 | 0/NA | 0/NA | 0/NA | 0/NA | Fig 4H | log10 gc/beta-actin | RT-qPCR |
| Achdout et al. [4] | 2021 | K18-hACE2 mice | female | H1N1 (A/Puerto Rico/8/1934, PR8) | SC2 (B.1) | 2nd | 2 | 10 | See table S1 | Lung | Tissue | 2 | 4.6** | NA | 4,56E+04 | 6,73E+04 | Fig 2B/D | Nasal Turbinates | Tissue | 2 | 11** | NA | 4,88E+04 | 7,26E+04 | Fig 2C/E | PFU/organ | Plaque-based |
| Achdout et al. [4] | 2021 | K18-hACE2 mice | female | H1N1 (A/Puerto Rico/8/1934, PR8) | SC2 (B.1) | 2nd | 2 | 10 | See table S1 | Lung | Tissue | 4 | 4,19E+08 | 3,15E+08 | 7,47E+04 | 1,50E+04 | Fig 2B/D | Nasal Turbinates | Tissue | 4 | 2.3** | NA | 1,64E+04 | 5,52E+04 | Fig 2C/E | PFU/organ | Plaque-based |
| Kinoshita et al. [5] | 2021 | Syrian hamster | female | H1N1 (A/Puerto Rico/8/1934, PR8) | SC2 (B.1.1) | simultaneous | 0 | 6 | See table S1 | Lung | Tissue | 4 | 7,00 | 6,90 | 11,00 | 11,00 | Table 1 | NA | NA | NA | NA | NA | NA | NA | NA | log10 gc/mg | RT-qPCR |
| Kinoshita et al. [5] | 2021 | Syrian hamster | female | H1N1 (A/Puerto Rico/8/1934, PR8) | SC2 (B.1.1) | simultaneous | 0 | 6 | See table S1 | Lung | Tissue | 7 | 3,70 | 2,70 | 7,30 | 7,10 | Table 1 | NA | NA | NA | NA | NA | NA | NA | NA | log10 gc/mg | RT-qPCR |
| Li et al. [6] | 2021 | hACE2 mice | female | H1N1 (A/Puerto Rico/8/1934, PR8) | SC2 (A) | 1st | -7 | 3 | See table S1 | Lung | Tissue | 2 | 12,66 | 12,14 | NA | NA | Fig 3D | Nasal | Tissue | 4 | 16,58 | 20,53 | NA | NA | Fig 3F | ct | RT-qPCR |
| Li et al. [6] | 2021 | hACE2 mice | female | H1N1 (A/Puerto Rico/8/1934, PR8) | SC2 (A) | 1st | -7 | 3 | See table S1 | Lung | Tissue | 4 | 13,37 | 13,99 | NA | NA | Fig 3D | Nasal | Swabs | 1 | 36,86 | 37,28 | NA | NA | Fig 3B | ct | RT-qPCR |
| Li et al. [6] | 2021 | hACE2 mice | female | H1N1 (A/Puerto Rico/8/1934, PR8) | SC2 (A) | 1st | -7 | 3 | See table S1 | Lung | Tissue | 7 | 17,91 | 21,60 | NA | NA | Fig 3D | Nasal | Swabs | 2 | 29,00 | 27,68 | NA | NA | Fig 3B | ct | RT-qPCR |
| Li et al. [6] | 2021 | hACE2 mice | female | H1N1 (A/Puerto Rico/8/1934, PR8) | SC2 (A) | 1st | -7 | 3 | See table S1 | Trachea | Tissue | 2 | 14,69 | 14,61 | NA | NA | Fig 3E | Nasal | Swabs | 3 | 21,18 | 23,81 | NA | NA | Fig 3B | ct | RT-qPCR |
| Li et al. [6] | 2021 | hACE2 mice | female | H1N1 (A/Puerto Rico/8/1934, PR8) | SC2 (A) | 1st | -7 | 3 | See table S1 | Trachea | Tissue | 4 | 20,16 | 21,90 | NA | NA | Fig 3E | Nasal | Swabs | 4 | 21,70 | 23,51 | NA | NA | Fig 3B | ct | RT-qPCR |
| Li et al. [6] | 2021 | hACE2 mice | female | H1N1 (A/Puerto Rico/8/1934, PR8) | SC2 (A) | 1st | -7 | 3 | See table S1 | Trachea | Tissue | 7 | 23,39 | 26,73 | NA | NA | Fig 3E | Nasal | Swabs | 5 | 20,06 | 22,18 | NA | NA | Fig 3B | ct | RT-qPCR |
| Li et al. [6] | 2021 | hACE2 mice | female | H1N1 (A/Puerto Rico/8/1934, PR8) | SC2 (A) | 1st | -7 | 3 | See table S1 | Pulmonary lymph node | Tissue | 4 | 22,17 | 26,62 | NA | NA | Fig 3F | Nasal | Swabs | 6 | 19,15 | 20,70 | NA | NA | Fig 3B | ct | RT-qPCR |
| Li et al. [6] | 2021 | hACE2 mice | female | H1N1 (A/Puerto Rico/8/1934, PR8) | SC2 (A) | 1st | -7 | 3 | See table S1 | NA | NA | NA | NA | NA | NA | NA | NA | Nasal | Swabs | 7 | 20,33 | 22,48 | NA | NA | Fig 3B | ct | RT-qPCR |
| Li et al. [6] | 2021 | hACE2 mice | female | H1N1 (A/Puerto Rico/8/1934, PR8) | SC2 (A) | 1st | -7 | 3 | See table S1 | NA | NA | NA | NA | NA | NA | NA | NA | Throat | Swabs | 1 | 35,77 | 35,85 | NA | NA | Fig 3C | ct | RT-qPCR |
| Li et al. [6] | 2021 | hACE2 mice | female | H1N1 (A/Puerto Rico/8/1934, PR8) | SC2 (A) | 1st | -7 | 3 | See table S1 | NA | NA | NA | NA | NA | NA | NA | NA | Throat | Swabs | 2 | 33,99 | 33,37 | NA | NA | Fig 3C | ct | RT-qPCR |
| Li et al. [6] | 2021 | hACE2 mice | female | H1N1 (A/Puerto Rico/8/1934, PR8) | SC2 (A) | 1st | -7 | 3 | See table S1 | NA | NA | NA | NA | NA | NA | NA | NA | Throat | Swabs | 3 | 28,74 | 28,87 | NA | NA | Fig 3C | ct | RT-qPCR |
| Li et al. [6] | 2021 | hACE2 mice | female | H1N1 (A/Puerto Rico/8/1934, PR8) | SC2 (A) | 1st | -7 | 3 | See table S1 | NA | NA | NA | NA | NA | NA | NA | NA | Throat | Swabs | 4 | 27,06 | 29,23 | NA | NA | Fig 3C | ct | RT-qPCR |
| Li et al. [6] | 2021 | hACE2 mice | female | H1N1 (A/Puerto Rico/8/1934, PR8) | SC2 (A) | 1st | -7 | 3 | See table S1 | NA | NA | NA | NA | NA | NA | NA | NA | Throat | Swabs | 5 | 26,99 | 28,55 | NA | NA | Fig 3C | ct | RT-qPCR |
| Li et al. [6] | 2021 | hACE2 mice | female | H1N1 (A/Puerto Rico/8/1934, PR8) | SC2 (A) | 1st | -7 | 3 | See table S1 | NA | NA | NA | NA | NA | NA | NA | NA | Throat | Swabs | 6 | 28,83 | 29,79 | NA | NA | Fig 3C | ct | RT-qPCR |
| Li et al. [6] | 2021 | hACE2 mice | female | H1N1 (A/Puerto Rico/8/1934, PR8) | SC2 (A) | 1st | -7 | 3 | See table S1 | NA | NA | NA | NA | NA | NA | NA | NA | Throat | Swabs | 7 | 31,28 | 33,06 | NA | NA | Fig 3C | ct | RT-qPCR |
| Li et al. [6] | 2021 | hACE2 mice | female | H1N1 (A/Puerto Rico/8/1934, PR8) | SC2 (A) | 1st | -14 | 3 | See table S1 | Lung | Tissue | 2 | 13,95 | 12,14 | NA | NA | Fig 3D | Nasal | Tissue | 4 | 17,30 | 20,53 | NA | NA | Fig 3F | ct | RT-qPCR |
| Li et al. [6] | 2021 | hACE2 mice | female | H1N1 (A/Puerto Rico/8/1934, PR8) | SC2 (A) | 1st | -14 | 3 | See table S1 | Lung | Tissue | 4 | 15,61 | 13,99 | NA | NA | Fig 3D | Nasal | Swabs | 1 | 34,82 | 37,28 | NA | NA | Fig 3B | ct | RT-qPCR |
| Li et al. [6] | 2021 | hACE2 mice | female | H1N1 (A/Puerto Rico/8/1934, PR8) | SC2 (A) | 1st | -14 | 3 | See table S1 | Lung | Tissue | 7 | 18,55 | 21,60 | NA | NA | Fig 3D | Nasal | Swabs | 2 | 26,41 | 27,68 | NA | NA | Fig 3B | ct | RT-qPCR |
| Li et al. [6] | 2021 | hACE2 mice | female | H1N1 (A/Puerto Rico/8/1934, PR8) | SC2 (A) | 1st | -14 | 3 | See table S1 | Trachea | Tissue | 2 | 14,75 | 14,61 | NA | NA | Fig 3E | Nasal | Swabs | 3 | 23,08 | 23,81 | NA | NA | Fig 3B | ct | RT-qPCR |
| Li et al. [6] | 2021 | hACE2 mice | female | H1N1 (A/Puerto Rico/8/1934, PR8) | SC2 (A) | 1st | -14 | 3 | See table S1 | Trachea | Tissue | 4 | 19,31 | 21,90 | NA | NA | Fig 3E | Nasal | Swabs | 4 | 21,29 | 23,51 | NA | NA | Fig 3B | ct | RT-qPCR |
| Li et al. [6] | 2021 | hACE2 mice | female | H1N1 (A/Puerto Rico/8/1934, PR8) | SC2 (A) | 1st | -14 | 3 | See table S1 | Trachea | Tissue | 7 | 24,31 | 26,73 | NA | NA | Fig 3E | Nasal | Swabs | 5 | 22,10 | 22,18 | NA | NA | Fig 3B | ct | RT-qPCR |
| Li et al. [6] | 2021 | hACE2 mice | female | H1N1 (A/Puerto Rico/8/1934, PR8) | SC2 (A) | 1st | -14 | 3 | See table S1 | Pulmonary lymph node | Tissue | NA | 19,92 | 26,62 | NA | NA | Fig 3F | Nasal | Swabs | 6 | 19,25 | 20,70 | NA | NA | Fig 3B | ct | RT-qPCR |
| Li et al. [6] | 2021 | hACE2 mice | female | H1N1 (A/Puerto Rico/8/1934, PR8) | SC2 (A) | 1st | -14 | 3 | See table S1 | NA | NA | NA | NA | NA | NA | NA | NA | Nasal | Swabs | 7 | 24,05 | 22,48 | NA | NA | Fig 3B | ct | RT-qPCR |
| Li et al. [6] | 2021 | hACE2 mice | female | H1N1 (A/Puerto Rico/8/1934, PR8) | SC2 (A) | 1st | -14 | 3 | See table S1 | NA | NA | NA | NA | NA | NA | NA | NA | Throat | Swabs | 1 | 35,42 | 35,85 | NA | NA | Fig 3C | ct | RT-qPCR |
| Li et al. [6] | 2021 | hACE2 mice | female | H1N1 (A/Puerto Rico/8/1934, PR8) | SC2 (A) | 1st | -14 | 3 | See table S1 | NA | NA | NA | NA | NA | NA | NA | NA | Throat | Swabs | 2 | 34,29 | 33,37 | NA | NA | Fig 3C | ct | RT-qPCR |
| Li et al. [6] | 2021 | hACE2 mice | female | H1N1 (A/Puerto Rico/8/1934, PR8) | SC2 (A) | 1st | -14 | 3 | See table S1 | NA | NA | NA | NA | NA | NA | NA | NA | Throat | Swabs | 3 | 28,78 | 28,87 | NA | NA | Fig 3C | ct | RT-qPCR |
| Li et al. [6] | 2021 | hACE2 mice | female | H1N1 (A/Puerto Rico/8/1934, PR8) | SC2 (A) | 1st | -14 | 3 | See table S1 | NA | NA | NA | NA | NA | NA | NA | NA | Throat | Swabs | 4 | 29,01 | 29,23 | NA | NA | Fig 3C | ct | RT-qPCR |
| Li et al. [6] | 2021 | hACE2 mice | female | H1N1 (A/Puerto Rico/8/1934, PR8) | SC2 (A) | 1st | -14 | 3 | See table S1 | NA | NA | NA | NA | NA | NA | NA | NA | Throat | Swabs | 5 | 28,87 | 28,55 | NA | NA | Fig 3C | ct | RT-qPCR |
| Li et al. [6] | 2021 | hACE2 mice | female | H1N1 (A/Puerto Rico/8/1934, PR8) | SC2 (A) | 1st | -14 | 3 | See table S1 | NA | NA | NA | NA | NA | NA | NA | NA | Throat | Swabs | 6 | 31,25 | 29,79 | NA | NA | Fig 3C | ct | RT-qPCR |
| Li et al. [6] | 2021 | hACE2 mice | female | H1N1 (A/Puerto Rico/8/1934, PR8) | SC2 (A) | 1st | -14 | 3 | See table S1 | NA | NA | NA | NA | NA | NA | NA | NA | Throat | Swabs | 7 | 30,52 | 33,06 | NA | NA | Fig 3C | ct | RT-qPCR |
| Halfmann et al. [7] | 2021 | Syrian hamster | female | H3N2 (A/Tokyo/UT-IMS3-1/2014) | SC2 (B.1) | simultaneous | 0 | 4 | See table S1 | Lung | Tissue | 3 | 3,47E+05 | 7,06E+05 | 8,73E+07 | 8,55E+07 | Fig 1A | Nasal Turbinates | Tissue | 3 | 1,89E+06 | 6,57E+05 | 7,23E+07 | 7,47E+07 | Fig 1A | PFU/g | Plaque-based |
| Halfmann et al. [7] | 2021 | Syrian hamster | female | H3N2 (A/Tokyo/UT-IMS3-1/2014) | SC2 (B.1) | 1st | -10 | 4 | See table S1 | Lung | Tissue | 3 | 2,74E+07 | 6,09E+07 | NA | NA | Fig 1B | Nasal Turbinates | Tissue | 3 | 3,26E+07 | 7,53E+07 | NA | NA | Fig 1B | PFU/g | Plaque-based |
| Halfmann et al. [7] | 2021 | Syrian hamster | female | H3N2 (A/Tokyo/UT-IMS3-1/2014) | SC2 (B.1) | 1st | -21 | 4 | See table S1 | Lung | Tissue | 3 | 3,40E+07 | 7,21E+07 | NA | NA | Fig 1C | Nasal Turbinates | Tissue | 3 | 4,79E+07 | 7,26E+07 | NA | NA | Fig 1C | PFU/g | Plaque-based |
| Halfmann et al. [7] | 2021 | Syrian hamster | female | H3N2 (A/Tokyo/UT-IMS3-1/2014) | SC2 (B.1) | 2nd | 10 | 4 | See table S1 | Lung | Tissue | 3 | 7,11E+07 | 6,72E+07 | NA | NA | Fig 1D | Nasal Turbinates | Tissue | 3 | NA | NA | 5,05E+07 | 4,36E+07 | Fig 1D | PFU/g | Plaque-based |
| Kim et al. [8] | 2022 | K18-hACE2 mice | female | H1N1 (A/California/04/2009) | SC2 (A) | 1st | -3 | 3 | See table S1 | Lung | Tissue | 2 | 3,73 | 5,94 | 4,47 | 4,66 | Fig 2B/C | NA | NA | NA | NA | NA | NA | NA | NA | log10 TCID50/g | Plaque-based |
| Kim et al. [8] | 2022 | K18-hACE2 mice | female | H1N1 (A/California/04/2009) | SC2 (A) | 1st | -3 | 3 | See table S1 | Lung | Tissue | 4 | 3,37 | 5,66 | 4,75 | 1,52 | Fig 2B/C | NA | NA | NA | NA | NA | NA | NA | NA | log10 TCID50/g | Plaque-based |
| Kim et al. [8] | 2022 | K18-hACE2 mice | female | H1N1 (A/California/04/2009) | SC2 (A) | 1st | -3 | 3 | See table S1 | Lung | Tissue | 7 | 3,72 | 1,52 | 1,52 | 1,52 | Fig 2B/C | NA | NA | NA | NA | NA | NA | NA | NA | log10 TCID50/g | Plaque-based |
| Kim et al. [8] | 2022 | K18-hACE2 mice | female | H1N1 (A/California/04/2009) | SC2 (A) | 2nd | 3 | 3 | See table S1 | Lung | Tissue | 2 | 4,39 | 5,92 | 3,71 | 4,65 | Fig 2A/D | NA | NA | NA | NA | NA | NA | NA | NA | log10 TCID50/g | Plaque-based |
| Kim et al. [8] | 2022 | K18-hACE2 mice | female | H1N1 (A/California/04/2009) | SC2 (A) | 2nd | 3 | 3 | See table S1 | Lung | Tissue | 4 | 3,09 | 5,63 | 2,15 | 1,52 | Fig 2A/D | NA | NA | NA | NA | NA | NA | NA | NA | log10 TCID50/g | Plaque-based |
| Kim et al. [8] | 2022 | K18-hACE2 mice | female | H1N1 (A/California/04/2009) | SC2 (A) | 2nd | 3 | 3 | See table S1 | Lung | Tissue | 7 | 2,78 | 1,52 | 1,52 | 1,52 | Fig 2A/D | NA | NA | NA | NA | NA | NA | NA | NA | log10 TCID50/g | Plaque-based |

| Author | Year | Animal model | Sex | IAV (strain) | SC2 (Pangolin lineage) | SC2 infection order | Days btwn infections | Sample size (per group) | Inoculation dose | Viral load in lower respiratory tract | | | | | | | Viral load in upper respiratory tract | | | | | | | Quantification Unit | Quantification Method |
|---|---|---|---|---|---|---|---|---|---|---|---|---|---|---|---|---|---|---|---|---|---|---|---|---|---|
| | | | | | | | | | | Tissue | Sample | Sampling day (dpi) | Coinf (IAV) | Monoinf (IAV) | Coinf (SC2) | Monoinf (SC2) | Figure/Table | Tissue | Sample | Sampling day (dpi) | Coinf (IAV) | Monoinf (IAV) | Coinf (SC2) | Monoinf (SC2) | Figure/Table | | |
| Huang et al. [9] | 2022 | Ferrets | female | H1N1 (A/California/07/2009) | SC2 (A) | simultaneous | 0 | 4 | See table S1 | Lung | Tissue | 4 | 4,80 | 5,35 | NA | NA | Figure 3A | Nasal | Swabs | 1 | 3,20 | 3,89 | NA | NA | Fig 2A-D | log10 PFU/mL | Plaque-based |
| Huang et al. [9] | 2022 | Ferrets | female | H1N1 (A/California/07/2009) | SC2 (A) | simultaneous | 0 | 4 | See table S1 | NA | NA | NA | NA | NA | NA | NA | NA | Nasal | Swabs | 3 | 3,06 | 4,42 | NA | NA | Fig 2A-D | log10 PFU/mL | Plaque-based |
| Huang et al. [9] | 2022 | Ferrets | female | H1N1 (A/California/07/2009) | SC2 (A) | simultaneous | 0 | 4 | See table S1 | NA | NA | NA | NA | NA | NA | NA | NA | Nasal | Swabs | 5 | 3,86 | 4,25 | NA | NA | Fig 2A-D | log10 PFU/mL | Plaque-based |
| Huang et al. [9] | 2022 | Ferrets | female | H1N1 (A/California/07/2009) | SC2 (A) | simultaneous | 0 | 4 | See table S1 | NA | NA | NA | NA | NA | NA | NA | NA | Nasal | Swabs | 7 | 1,00 | 1,00 | NA | NA | Fig 2A-D | log10 PFU/mL | Plaque-based |
| Huang et al. [9] | 2022 | Ferrets | female | H1N1 (A/California/07/2009) | SC2 (A) | simultaneous | 0 | 4 | See table S1 | NA | NA | NA | NA | NA | NA | NA | NA | Nasal | Swabs | 9 | 1,00 | 1,00 | NA | NA | Fig 2A-D | log10 PFU/mL | Plaque-based |
| Huang et al. [9] | 2022 | Ferrets | female | H3N2 (A/Kansas/14/2017) | SC2 (A) | simultaneous | 0 | 4 | See table S1 | Lung | Tissue | 4 | 1,00 | 1,00 | NA | NA | Figure 3A | Nasal | Swabs | 1 | 2,29 | 3,28 | NA | NA | Fig 2A-D | log10 PFU/mL | Plaque-based |
| Huang et al. [9] | 2022 | Ferrets | female | H3N2 (A/Kansas/14/2017) | SC2 (A) | simultaneous | 0 | 4 | See table S1 | NA | NA | NA | NA | NA | NA | NA | NA | Nasal | Swabs | 3 | 2,82 | 3,16 | NA | NA | Fig 2A-D | log10 PFU/mL | Plaque-based |
| Huang et al. [9] | 2022 | Ferrets | female | H3N2 (A/Kansas/14/2017) | SC2 (A) | simultaneous | 0 | 4 | See table S1 | NA | NA | NA | NA | NA | NA | NA | NA | Nasal | Swabs | 5 | 1,93 | 2,32 | NA | NA | Fig 2A-D | log10 PFU/mL | Plaque-based |
| Huang et al. [9] | 2022 | Ferrets | female | H3N2 (A/Kansas/14/2017) | SC2 (A) | simultaneous | 0 | 4 | See table S1 | NA | NA | NA | NA | NA | NA | NA | NA | Nasal | Swabs | 7 | 1,00 | 1,00 | NA | NA | Fig 2A-D | log10 PFU/mL | Plaque-based |
| Huang et al. [9] | 2022 | Ferrets | female | H3N2 (A/Kansas/14/2017) | SC2 (A) | simultaneous | 0 | 4 | See table S1 | NA | NA | NA | NA | NA | NA | NA | NA | Nasal | Swabs | 9 | 1,00 | 1,00 | NA | NA | Fig 2A-D | log10 PFU/mL | Plaque-based |
| Huang et al. [9] | 2022 | Ferrets | female | H1N1 (A/California/07/2009) | SC2 (A) | simultaneous | 0 | 4 | See table S1 | Lung | Tissue | 4 | NA | NA | 4,30 | 4,30 | Figure 3B | Nasal | Swabs | 1 | NA | NA | 6,50 | 5,25 | Fig 2E-G | log2 TCID50/mL | Median tissue culture infectious dose |
| Huang et al. [9] | 2022 | Ferrets | female | H1N1 (A/California/07/2009) | SC2 (A) | simultaneous | 0 | 4 | See table S1 | NA | NA | NA | NA | NA | NA | NA | NA | Nasal | Swabs | 3 | NA | NA | 4,30 | 6,97 | Fig 2E-G | log2 TCID50/mL | Median tissue culture infectious dose |
| Huang et al. [9] | 2022 | Ferrets | female | H1N1 (A/California/07/2009) | SC2 (A) | simultaneous | 0 | 4 | See table S1 | NA | NA | NA | NA | NA | NA | NA | NA | Nasal | Swabs | 5 | NA | NA | 4,30 | 5,87 | Fig 2E-G | log2 TCID50/mL | Median tissue culture infectious dose |
| Huang et al. [9] | 2022 | Ferrets | female | H1N1 (A/California/07/2009) | SC2 (A) | simultaneous | 0 | 4 | See table S1 | NA | NA | NA | NA | NA | NA | NA | NA | Nasal | Swabs | 7 | NA | NA | 4,30 | 4,56 | Fig 2E-G | log2 TCID50/mL | Median tissue culture infectious dose |
| Huang et al. [9] | 2022 | Ferrets | female | H1N1 (A/California/07/2009) | SC2 (A) | simultaneous | 0 | 4 | See table S1 | NA | NA | NA | NA | NA | NA | NA | NA | Nasal | Swabs | 9 | NA | NA | 4,30 | 4,55 | Fig 2E-G | log2 TCID50/mL | Median tissue culture infectious dose |
| Huang et al. [9] | 2022 | Ferrets | female | H3N2 (A/Kansas/14/2017) | SC2 (A) | simultaneous | 0 | 4 | See table S1 | Lung | Tissue | 4 | NA | NA | 4,30 | 4,30 | Figure 3B | Nasal | Swabs | 1 | NA | NA | 6,88 | 5,25 | Fig 2E-G | log2 TCID50/mL | Median tissue culture infectious dose |
| Huang et al. [9] | 2022 | Ferrets | female | H3N2 (A/Kansas/14/2017) | SC2 (A) | simultaneous | 0 | 4 | See table S1 | NA | NA | NA | NA | NA | NA | NA | NA | Nasal | Swabs | 3 | NA | NA | 4,30 | 6,97 | Fig 2E-G | log2 TCID50/mL | Median tissue culture infectious dose |
| Huang et al. [9] | 2022 | Ferrets | female | H3N2 (A/Kansas/14/2017) | SC2 (A) | simultaneous | 0 | 4 | See table S1 | NA | NA | NA | NA | NA | NA | NA | NA | Nasal | Swabs | 5 | NA | NA | 4,30 | 5,87 | Fig 2E-G | log2 TCID50/mL | Median tissue culture infectious dose |
| Huang et al. [9] | 2022 | Ferrets | female | H3N2 (A/Kansas/14/2017) | SC2 (A) | simultaneous | 0 | 4 | See table S1 | NA | NA | NA | NA | NA | NA | NA | NA | Nasal | Swabs | 7 | NA | NA | 4,30 | 4,56 | Fig 2E-G | log2 TCID50/mL | Median tissue culture infectious dose |
| Huang et al. [9] | 2022 | Ferrets | female | H3N2 (A/Kansas/14/2017) | SC2 (A) | simultaneous | 0 | 4 | See table S1 | NA | NA | NA | NA | NA | NA | NA | NA | Nasal | Swabs | 9 | NA | NA | 4,30 | 4,55 | Fig 2E-G | log2 TCID50/mL | Median tissue culture infectious dose |
| Huang et al. [9] | 2022 | Ferrets | female | H1N1 (A/California/07/2009) | SC2 (A) | simultaneous | 0 | 4 | See table S1 | NA | NA | NA | NA | NA | NA | NA | NA | Nasal | Swabs | 1 | NA | NA | 3,30 | 3,16 | Fig 8A/C | log10 gc/uL | RT-qPCR |
| Huang et al. [9] | 2022 | Ferrets | female | H1N1 (A/California/07/2009) | SC2 (A) | simultaneous | 0 | 4 | See table S1 | NA | NA | NA | NA | NA | NA | NA | NA | Nasal | Swabs | 3 | NA | NA | 2,18 | 4,64 | Fig 8A/C | log10 gc/uL | RT-qPCR |
| Huang et al. [9] | 2022 | Ferrets | female | H1N1 (A/California/07/2009) | SC2 (A) | simultaneous | 0 | 4 | See table S1 | NA | NA | NA | NA | NA | NA | NA | NA | Nasal | Swabs | 5 | NA | NA | 1,33 | 3,74 | Fig 8A/C | log10 gc/uL | RT-qPCR |
| Huang et al. [9] | 2022 | Ferrets | female | H1N1 (A/California/07/2009) | SC2 (A) | simultaneous | 0 | 4 | See table S1 | NA | NA | NA | NA | NA | NA | NA | NA | Nasal | Swabs | 7 | NA | NA | 1,51 | 3,40 | Fig 8A/C | log10 gc/uL | RT-qPCR |
| Huang et al. [9] | 2022 | Ferrets | female | H1N1 (A/California/07/2009) | SC2 (A) | simultaneous | 0 | 4 | See table S1 | NA | NA | NA | NA | NA | NA | NA | NA | Nasal | Swabs | 9 | NA | NA | 0,48 | 0,00 | Fig 8A/C | log10 gc/uL | RT-qPCR |
| Huang et al. [9] | 2022 | Ferrets | female | H3N2 (A/Kansas/14/2017) | SC2 (A) | simultaneous | 0 | 4 | See table S1 | NA | NA | NA | NA | NA | NA | NA | NA | Nasal | Swabs | 1 | NA | NA | 3,16 | 3,16 | Fig 8B/C | log10 gc/uL | RT-qPCR |
| Huang et al. [9] | 2022 | Ferrets | female | H3N2 (A/Kansas/14/2017) | SC2 (A) | simultaneous | 0 | 4 | See table S1 | NA | NA | NA | NA | NA | NA | NA | NA | Nasal | Swabs | 3 | NA | NA | 2,19 | 4,64 | Fig 8B/C | log10 gc/uL | RT-qPCR |
| Huang et al. [9] | 2022 | Ferrets | female | H3N2 (A/Kansas/14/2017) | SC2 (A) | simultaneous | 0 | 4 | See table S1 | NA | NA | NA | NA | NA | NA | NA | NA | Nasal | Swabs | 5 | NA | NA | 0,54 | 3,74 | Fig 8B/C | log10 gc/uL | RT-qPCR |
| Huang et al. [9] | 2022 | Ferrets | female | H3N2 (A/Kansas/14/2017) | SC2 (A) | simultaneous | 0 | 4 | See table S1 | NA | NA | NA | NA | NA | NA | NA | NA | Nasal | Swabs | 7 | NA | NA | 0,00 | 3,40 | Fig 8B/C | log10 gc/uL | RT-qPCR |
| Huang et al. [9] | 2022 | Ferrets | female | H3N2 (A/Kansas/14/2017) | SC2 (A) | simultaneous | 0 | 4 | See table S1 | NA | NA | NA | NA | NA | NA | NA | NA | Nasal | Swabs | 9 | NA | NA | 0,00 | 0,00 | Fig 8B/C | log10 gc/uL | RT-qPCR |
| Kim et al. [10] | 2022 | Syrian hamster | male | H1N1 (A/California/04/2009) | SC2 (B) | 1st | -1 | Unreported | See table S1 | Lung | Tissue | 1 | 1,93 | 1,88 | 5,48 | 4,15 | Fig 1E/H | Nasal Turbinates | Tissue | 1 | 1,91 | 2,99 | 5,70 | 5,80 | Fig 1C/F | log10 TCID50/mL | RT-qPCR |
| Kim et al. [10] | 2022 | Syrian hamster | male | H1N1 (A/California/04/2009) | SC2 (B) | 1st | -1 | Unreported | See table S1 | Lung | Tissue | 3 | 2,68 | 1,00 | 4,68 | 3,93 | Fig 1E/H | Nasal Turbinates | Tissue | 3 | 2,42 | 3,66 | 4,55 | 4,61 | Fig 1C/F | log10 TCID50/mL | RT-qPCR |
| Kim et al. [10] | 2022 | Syrian hamster | male | H1N1 (A/California/04/2009) | SC2 (B) | 1st | -1 | Unreported | See table S1 | Lung | Tissue | 6 | 1,00 | 1,00 | 2,81 | 2,43 | Fig 1E/H | Nasal Turbinates | Tissue | 6 | 3,64 | 1,00 | 3,62 | 4,00 | Fig 1C/F | log10 TCID50/mL | RT-qPCR |
| Kim et al. [10] | 2022 | Syrian hamster | male | H1N1 (A/California/04/2009) | SC2 (B) | 1st | -1 | Unreported | See table S1 | Trachea | Tissue | 1 | 1,93 | 1,88 | 3,23 | 2,63 | Fig 1D/G | NA | NA | NA | NA | NA | NA | NA | NA | log10 TCID50/mL | RT-qPCR |
| Kim et al. [10] | 2022 | Syrian hamster | male | H1N1 (A/California/04/2009) | SC2 (B) | 1st | -1 | Unreported | See table S1 | Trachea | Tissue | 3 | 2,50 | 1,00 | 3,86 | 1,99 | Fig 1D/G | NA | NA | NA | NA | NA | NA | NA | NA | log10 TCID50/mL | RT-qPCR |
| Kim et al. [10] | 2022 | Syrian hamster | male | H1N1 (A/California/04/2009) | SC2 (B) | 1st | -1 | Unreported | See table S1 | Trachea | Tissue | 6 | 1,00 | 1,00 | 1,73 | 1,00 | Fig 1D/G | NA | NA | NA | NA | NA | NA | NA | NA | log10 TCID50/mL | RT-qPCR |
| Kim et al. [10] | 2022 | Syrian hamster | male | H1N1 (A/California/04/2009) | SC2 (B) | 2nd | 1 | Unreported | See table S1 | Lung | Tissue | 1 | 1,27 | 1,88 | 4,42 | 4,15 | Fig 1E/H | Nasal Turbinates | Tissue | 1 | 2,52 | 2,99 | 5,82 | 5,80 | Fig 1C/F | log10 TCID50/mL | RT-qPCR |
| Kim et al. [10] | 2022 | Syrian hamster | male | H1N1 (A/California/04/2009) | SC2 (B) | 2nd | 1 | Unreported | See table S1 | Lung | Tissue | 3 | 2,32 | 1,00 | 4,34 | 3,93 | Fig 1E/H | Nasal Turbinates | Tissue | 3 | 2,17 | 3,66 | 4,39 | 4,61 | Fig 1C/F | log10 TCID50/mL | RT-qPCR |
| Kim et al. [10] | 2022 | Syrian hamster | male | H1N1 (A/California/04/2009) | SC2 (B) | 2nd | 1 | Unreported | See table S1 | Lung | Tissue | 6 | 1,00 | 1,00 | 2,39 | 2,43 | Fig 1E/H | Nasal Turbinates | Tissue | 6 | 1,98 | 1,00 | 3,65 | 4,00 | Fig 1C/F | log10 TCID50/mL | RT-qPCR |
| Kim et al. [10] | 2022 | Syrian hamster | male | H1N1 (A/California/04/2009) | SC2 (B) | 2nd | 1 | Unreported | See table S1 | Trachea | Tissue | 1 | 1,30 | 1,88 | 1,00 | 2,63 | Fig 1D/G | NA | NA | NA | NA | NA | NA | NA | NA | log10 TCID50/mL | RT-qPCR |
| Kim et al. [10] | 2022 | Syrian hamster | male | H1N1 (A/California/04/2009) | SC2 (B) | 2nd | 1 | Unreported | See table S1 | Trachea | Tissue | 3 | 2,32 | 1,00 | 1,00 | 1,99 | Fig 1D/G | NA | NA | NA | NA | NA | NA | NA | NA | log10 TCID50/mL | RT-qPCR |
| Kim et al. [10] | 2022 | Syrian hamster | male | H1N1 (A/California/04/2009) | SC2 (B) | 2nd | 1 | Unreported | See table S1 | Trachea | Tissue | 6 | 1,00 | 1,00 | 1,00 | 1,00 | Fig 1D/G | NA | NA | NA | NA | NA | NA | NA | NA | log10 TCID50/mL | RT-qPCR |
| Oishi et al. [11] | 2022 | Syrian hamster | male | H1N1 (A/California/04/2009) | SC2 (A) | simultaneous | 0 | 4 | See table S1 | Lung | Tissue | 1 | 5,74 | 5,66 | 7,17 | 7,17 | Fig 2B/C | NA | NA | NA | NA | NA | NA | NA | NA | log10 PFU/g | Plaque-based |
| Oishi et al. [11] | 2022 | Syrian hamster | male | H1N1 (A/California/04/2009) | SC2 (A) | simultaneous | 0 | 4 | See table S1 | Lung | Tissue | 3 | 6,90 | 7,04 | 6,56 | 8,07 | Fig 2B/C | NA | NA | NA | NA | NA | NA | NA | NA | log10 PFU/g | Plaque-based |
| Oishi et al. [11] | 2022 | Syrian hamster | male | H1N1 (A/California/04/2009) | SC2 (A) | simultaneous | 0 | 4 | See table S1 | Lung | Tissue | 5 | 3,32 | 3,32 | 3,12 | 6,92 | Fig 2B/C | NA | NA | NA | NA | NA | NA | NA | NA | log10 PFU/g | Plaque-based |
| Oishi et al. [11] | 2022 | Syrian hamster | male | H1N1 (A/California/04/2009) | SC2 (A) | simultaneous | 0 | 4 | See table S1 | Lung | Tissue | 7 | 3,00 | 3,11 | 3,00 | 3,12 | Fig 2B/C | NA | NA | NA | NA | NA | NA | NA | NA | log10 PFU/g | Plaque-based |
| Oishi et al. [11] | 2022 | Syrian hamster | male | H1N1 (A/California/04/2009) | SC2 (A) | simultaneous | 0 | 4 | See table S1 | Lung | Tissue | 14 | 3,00 | 3,00 | 3,00 | 3,00 | Fig 2B/C | NA | NA | NA | NA | NA | NA | NA | NA | log10 PFU/g | Plaque-based |
| Oishi et al. [11] | 2022 | Syrian hamster | male | H1N1 (A/California/04/2009) | SC2 (A) | 1st | -3 | 3 | See table S1 | Lung | Tissue | 1 | 4,85 | 5,49 | 7,39 | 7,37 | Fig 3E/F | NA | NA | NA | NA | NA | NA | NA | NA | log10 PFU/g | Plaque-based |
| Oishi et al. [11] | 2022 | Syrian hamster | male | H1N1 (A/California/04/2009) | SC2 (A) | 1st | -3 | 3 | See table S1 | Lung | Tissue | 3 | 5,92 | 6,59 | 3,12 | 3,11 | Fig 3E/F | NA | NA | NA | NA | NA | NA | NA | NA | log10 PFU/g | Plaque-based |
| Oishi et al. [11] | 2022 | Syrian hamster | male | H1N1 (A/California/04/2009) | SC2 (A) | 1st | -3 | 3 | See table S1 | Lung | Tissue | 5 | 3,92 | 3,88 | 3,00 | 3,00 | Fig 3E/F | NA | NA | NA | NA | NA | NA | NA | NA | log10 PFU/g | Plaque-based |
| Oishi et al. [11] | 2022 | Syrian hamster | male | H1N1 (A/California/04/2009) | SC2 (A) | 2nd | 3 | 3 | See table S1 | Lung | Tissue | 1 | 7,70 | 7,24 | 3,23 | 7,08 | Fig 3B/C | NA | NA | NA | NA | NA | NA | NA | NA | log10 PFU/g | Plaque-based |
| Oishi et al. [11] | 2022 | Syrian hamster | male | H1N1 (A/California/04/2009) | SC2 (A) | 2nd | 3 | 3 | See table S1 | Lung | Tissue | 3 | 3,00 | 3,00 | 7,49 | 9,09 | Fig 3B/C | NA | NA | NA | NA | NA | NA | NA | NA | log10 PFU/g | Plaque-based |
| Oishi et al. [11] | 2022 | Syrian hamster | male | H1N1 (A/California/04/2009) | SC2 (A) | 2nd | 3 | 3 | See table S1 | Lung | Tissue | 5 | 3,00 | 3,00 | 3,39 | 7,31 | Fig 3B/C | NA | NA | NA | NA | NA | NA | NA | NA | log10 PFU/g | Plaque-based |
| Oishi et al. [11] | 2022 | Syrian hamster | male | H1N1 (A/California/04/2009) | SC2 (A) | 2nd | 7 | 4 | See table S1 | Lung | Tissue | 1 | 3,00 | 3,00 | 3,00 | 6,79 | Fig 4B | NA | NA | NA | NA | NA | NA | NA | NA | log10 PFU/g | Plaque-based |
| Oishi et al. [11] | 2022 | Syrian hamster | male | H1N1 (A/California/04/2009) | SC2 (A) | 2nd | 7 | 4 | See table S1 | Lung | Tissue | 3 | 3,00 | 3,00 | 7,69 | 8,93 | Fig 4B | NA | NA | NA | NA | NA | NA | NA | NA | log10 PFU/g | Plaque-based |
| Oishi et al. [11] | 2022 | Syrian hamster | male | H1N1 (A/California/04/2009) | SC2 (A) | 2nd | 7 | 4 | See table S1 | Lung | Tissue | 5 | 3,00 | 3,00 | 7,15 | 7,44 | Fig 4B | NA | NA | NA | NA | NA | NA | NA | NA | log10 PFU/g | Plaque-based |
| Oishi et al. [11] | 2022 | Syrian hamster | male | H1N1 (A/California/04/2009) | SC2 (A) | 2nd | 7 | 4 | See table S1 | Lung | Tissue | 7 | 3,00 | 3,00 | 3,00 | 3,00 | Fig 4B | NA | NA | NA | NA | NA | NA | NA | NA | log10 PFU/g | Plaque-based |
| Oishi et al. [11] | 2022 | Syrian hamster | male | H1N1 (A/California/04/2009) | SC2 (A) | 2nd | 14 | 4 | See table S1 | Lung | Tissue | 1 | 3,00 | 3,00 | 6,76 | 8,02 | Fig 4D | NA | NA | NA | NA | NA | NA | NA | NA | log10 PFU/g | Plaque-based |
| Oishi et al. [11] | 2022 | Syrian hamster | male | H1N1 (A/California/04/2009) | SC2 (A) | 2nd | 14 | 4 | See table S1 | Lung | Tissue | 3 | 3,00 | 3,00 | 8,15 | 8,61 | Fig 4D | NA | NA | NA | NA | NA | NA | NA | NA | log10 PFU/g | Plaque-based |
| Oishi et al. [11] | 2022 | Syrian hamster | male | H1N1 (A/California/04/2009) | SC2 (A) | 2nd | 14 | 4 | See table S1 | Lung | Tissue | 5 | 3,00 | 3,00 | 7,03 | 7,55 | Fig 4D | NA | NA | NA | NA | NA | NA | NA | NA | log10 PFU/g | Plaque-based |
| Oishi et al. [11] | 2022 | Syrian hamster | male | H1N1 (A/California/04/2009) | SC2 (A) | 2nd | 14 | 4 | See table S1 | Lung | Tissue | 7 | 3,00 | 3,00 | 3,00 | 3,00 | Fig 4D | NA | NA | NA | NA | NA | NA | NA | NA | log10 PFU/g | Plaque-based |

**Abbreviations** K18-hACE2 mice: transgenic mice expressing human angiotensin-converting enzyme 2 (hACE2) controlled by the human cytokeratin 18 promoter, dpi: days post infection, PFU: plaque-forming unit, CCID50: Cell culture infectious dose 50%, TCID50: Tissue culture infectious dose 50%, ct: cycle threshold, gc: gene copies, GAPDH: glyceraldehyde 3-phosphate dehydrogenase, RT-qPCR: reverse transcription quantitative

**Remark** Viral load ratio (co-infection:mono-infection) reported directly in the articles.

**Appendix 3: Table S3. Observational studies examining the association between pneumococcal vaccination history and COVID-19 [1–4].**

| Author | Design | Study population | Method to control for confounding | Outcome | Exposure | *Effect Estimate (95% CI) |
|---|---|---|---|---|---|---|
| Lewnard et al. [1] | Cohort-Perspective | Adults ≥65yrs | Using doubly robust inverse propensity weighting in Cox model, corrected for negative control (zoster vaccine recipients) | COVID-19 diagnosis | PCV PPSV | HR=0.65 (0.59, 0.72) HR=1.19 (1.05, 1.36) |
| | | | | COVID-19 hospitalization | | HR=0.68 (0.57, 0.83) HR=1.02 (0.78, 1.29) |
| | | | | COVID-19 mortality | | HR=0.68 (0.49, 0.95) HR=1.28 (0.77, 2.01) |
| Patwardhan et al. [2] | Cohort-Retrospective | Children <20yrs PCR+ve for SARS-CoV-2 | Adjusted for race, sex, age, month of diagnosis, comorbidity, allergy/asthma, obesity, smoke exposure | Subjective and objective symptom | Flu vac PCV | OR=0.714 (0.529, 0.964) OR=0.482(0.277,0.837) |
| | | | | Respiratory symptom | | OR=0.672(0.500, 0.903) OR=0.412(0.234, 0.725) |
| | | | | Severity (objective symptom-ve vs objective symptom+ve) | | OR=0.678(0.482,0.934) OR=0.765(00.428,1.368) |
| Rivas et al. [3] | Cohort-Retrospective | HCW | Adjusted for sex, age | **Blood test: anti–SARS-CoV-2 IgG+ index value >0.4 | BCG vac Men vac PPSV Flu vac | OR=0.76 (0.57, 0.99) OR=0.90 (0.69, 1.17) OR=0.99 (0.71, 1.36) OR=1.84 (0.57, 11.27) |

| Author | Design | Study population | Method to control for confounding | Outcome | Exposure | *Effect Estimate (95% CI) |
|---|---|---|---|---|---|---|
| Fernández-Prada et al. [4] | Case-control | Suspected cases (≥1 epi criterion + ≥1 clinical criterion) | Controls matched to cases based on sex, age, severity (hospital/home) | PCR+ve for SARS-CoV-2 | Flu vac PCV PPSV | OR=1.7 (0.957-3.254) OR=0.4 (0.170-1.006) OR=0.7 (0.284-2.097) |

**Abbreviations** PCV: pneumococcal conjugate vaccine, PPSV: pneumococcal polysaccharide vaccine, Flu vac: inluenza vaccine, BCG vac: Bacillus Calmette–Guérin vaccine, Men vac: Meningococcal vaccine, HCW: health care workers, PCR: Polymerase chain reaction, CI: confidence intervals, HR: hazard ratio, OR: odds ratio.
*Remark 1 Effect estimates were HR or OR directly extracted from studies.
**Remark 2 The presence of anti–SARS-CoV-2 IgG+ was interpreted as prior SARS-CoV-2 infection in the context before large-scale COVID-vaccine campaigns were implemented.

**Appendix 4: Table S4. Observational studies examining the association between prior respiratory infections and COVID-19 [1–4].**

| Author | Design | Study population | Method to control for confounding | Outcome | Exposure | *Effect Estimate (95% CI or SE) |
|---|---|---|---|---|---|---|
| Kim et al. [1] | Case-control | Individuals covered by national health insurance | Controls matched to cases based on sex, age, income; model adjusted for CCI scores, asthma, COPD, hypertension | COVID-19 diagnosis | Measured in rolling window prior to outcome: a) 1-14 days b) 1-30 days c) 1-90 days  Prescription of antiviral for influenza treatment  URI (ICD-10 J00 to J06) | a) OR=3.07 (1.16, 5.85) b) OR=1.18 (0.72, 1.91) c) OR=1.91 (1.54, 2.37) a) OR=6.95 (6.38, 7.58) b) OR=4.99 (4.64, 5.37) c) OR=2.70 (2.55, 2.86) |
| | | | | COVID-19 morbidity | | a) OR=3.64 (1.55, 9.21) b) OR=3.59 (1.42, 9.05) c) OR=1.54 (0.84, 2.84) a) OR=1.40 (1.11, 1.78) b) OR=1.28 (1.02, 1.61) c) OR=1.17 (0.95, 1.43) |
| | | | | COVID-19 mortality | | a) OR=3.66 (0.71, 18.81) b) OR=3.12 (0.64, 15.19) c) OR=1.62 (0.52, 4.47) a) OR= 0.90 (0.58, 1.40) b) OR= 0.82 (0.54, 1.26) c) OR= 0.77 (0.54, 1.10) |

| Author | Design | Study population | Method to control for confounding | Outcome | Exposure | *Effect Estimate (95% CI or SE) |
|---|---|---|---|---|---|---|
| Sager et al. [2] | Cohort-Retrospective | Adults>18 whose SARS-CoV-2 PCR result documented >=7 days after hCoV PCR result | Adjusted for race, COPD, HIV, number of comorbidities, level of clinical care (†or age, sex, BMI, DM). | PCR+ve for SARS-CoV-2 (among tested) | PCR+ve for hCoV | OR=0.9 (0.6–1.4) |
| | | | | COVID-19 hospitalization (among SARS-CoV-2 +ve) | | OR=1.6 (0.8–3.2) |
| | | | | COVID-19 ICU admission (among hospitalized) | | OR=0.1 (0.0–0.7) †OR=0.1 (0.1–0.9) |
| | | | | COVID-19 MV (among hospitalized) | | OR=0.0 (0.0–1.0) |
| Anderson et al. [3] | Case-control | Individuals whose serum sample were collected before pandemic (2020-03) | Controls matched to cases based on sex, age, race. | PCR+ve for SARS-CoV-2 (among tested) | hCoV antibodies (OC43 Spike Titer) | **$\beta = 1 \times 10^{-6}$ (SE: $2 \times 10^{-5}$) |
| | | | | COVID-19 hospitalization (among SARS-CoV-2 +ve) | | **$\beta = 1 \times 10^{-5}$ (SE: $3 \times 10^{-5}$) |
| | | | | COVID-19 severe hospitalization (among SARS-CoV-2 +ve) | | **$\beta = 2 \times 10^{-5}$ (SE: $5 \times 10^{-5}$) |

| Author | Design | Study population | Method to control for confounding | Outcome | Exposure | *Effect Estimate (95% CI or SE) |
|---|---|---|---|---|---|---|
| Aran et al. [4] | Case-control | Individuals tested for SARS-CoV-2 with >=12mo enrollment in a private insurance prior to test | Adjusted for sex, age, health-seeking behavior. | PCR+ve for SARS-CoV-2 | Measured within 1 year before study: URI (ICD-10 J01, J02.8/9, J20.9) | OR=0.76 (0.75, 0.77) |

**Abbreviations** CCI: Charlson Comorbidity Index, COPD: chronic obstructive pulmonary disease, URI: upper respiratory infections, HIV: human immunodeficiency virus, BMI: Body Mass Index, DM: Diabetes mellitus, ICU: intensive care unit, MV: mechanical ventilation, ICD: International Classification of Diseases, PCR: Polymerase chain reaction, CI: confidence intervals, OR: odds ratio, SE: standard error.
*Remark 1 Effect estimates were OR or β (regression coefficient) directly extracted from studies.
**Remark 2 β, the regression coefficient, can be interpreted as the increase in ln(odds) for COVID-19 outcome per unit increase in exposure.

**Appendix 5: Model details**

We developed deterministic compartment models of the two interacting pathogens. Following Shrestha et al. we used a double index notation, e.g $X_{Y,Z}$ where Y gives the state of pathogen 1 and Z the state of pathogen 2 [1].

Bacteria-virus interaction model:

The bacteria-virus model was constructed such that pathogen 1 is the bacteria and pathogen 2 is the virus. We assumed the interaction was asymmetric, such that colonization with bacteria impacts transmission of the virus, but infection with the virus has no impact on the bacterial dynamics. The model was defined by 2 x 3 = 6 ordinary differential equations as represented in Figure S1, where the disease states are {S, C} for bacteria, and {S, I, R} for the virus.

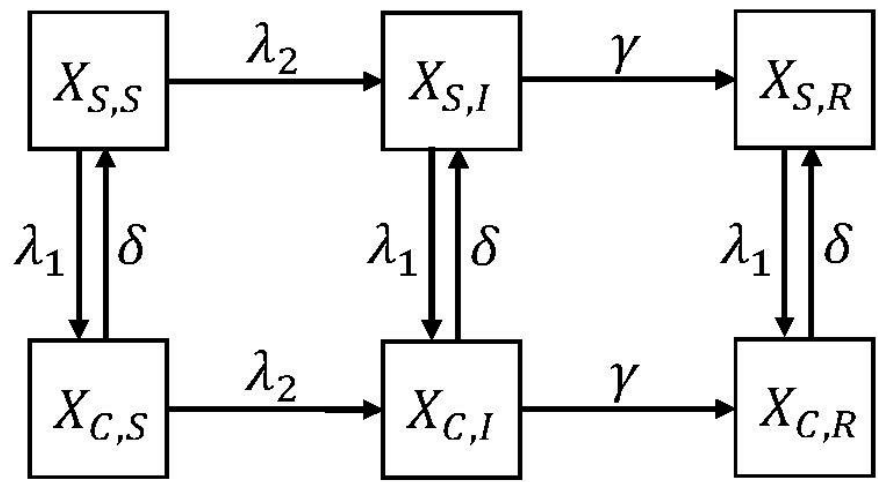

**Figure S1: Schematic of the bacteria-virus interaction model.** Horizontal transitions between compartments are due to virus dynamics, vertical transitions are due to bacteria dynamics.

The forces of infection ($\lambda$) and prevalences ($p$) for each disease are defined as:

$$\lambda_1(t) = (\delta/(1 - C*))p_1(t)$$

$$p_1(t) = X_{C,S} + X_{C,I} + X_{C,R}$$

$$\lambda_2(t) = R_0\,\gamma\,p_2(t)$$

$$p_2(t) = X_{S,I} + \theta\,X_{C,I}.$$

Where $1/\delta$ is the average period of bacteria colonization, $C^* = \lim_{t\to\infty}(X_{C,S} + X_{C,I} + X_{C,R})$ is the endemic colonization prevalence. $R_0$ is the initial reproduction number of the virus, and $1/\gamma$ is the average recovery period of the virus. $\theta$ is an interaction parameter indicating the impact of bacterial carriage on transmission.

As an example we consider the *S. pneumoniae* - SARS-CoV-2 interacting system, hence assuming the bacteria is *S. pneumoniae* and the virus is SARS-CoV-2. Parameter values are detailed in Table S5. The model was run for 365 days and the peak viral incidence was calculated for varying rate of bacterial colonization and varying transmission interaction parameter. The model was implemented in the R [2] packages 'pomp' [3], and 'tidyverse' [4]. Plots were created with 'ggplot2' [5], 'patchwork' [6], 'scico' [7] and 'Microsoft PowerPoint'. All code is available at https://github.com/egoult/pathogen_coinfections .

**Table S5. Parameters used for *S. pneumoniae* - SARS-CoV-2 interaction model.**

| Parameter | Meaning | Fixed values | Source |
|---|---|---|---|
| $C^*$ | Bacterial colonization prevalence | 0–0.6 | [8,9] |

| Parameter | Meaning | Fixed values | Source |
|---|---|---|---|
| $1/\delta$ | Duration of bacterial colonization | 50 days | [10,11] |
| $I_0$ | Initial fraction infected with virus | $1 \times 10^{-5}$ | Assumption |
| $R_0$ | Virus basic reproductive number | 2 | [12] |
| $1/\gamma$ | Virus recovery period | 9 days | [12,13] |
| $\theta$ | Coinfection impact on viral transmission | 0.2, 0.5, 0.8, 1.0, 1.2, 2.0, 5.0 | Assumption |

Virus-virus interaction model

The virus virus interaction model was also constructed asymmetrically, so infection with virus 1 impacts transmission of virus 2, but infection with virus 2 has no impact on virus 1. The model is defined in 4 x 4 = 16 ordinary differential equations as shown in Figure S2 with each virus having the disease states {SEIR} [14].

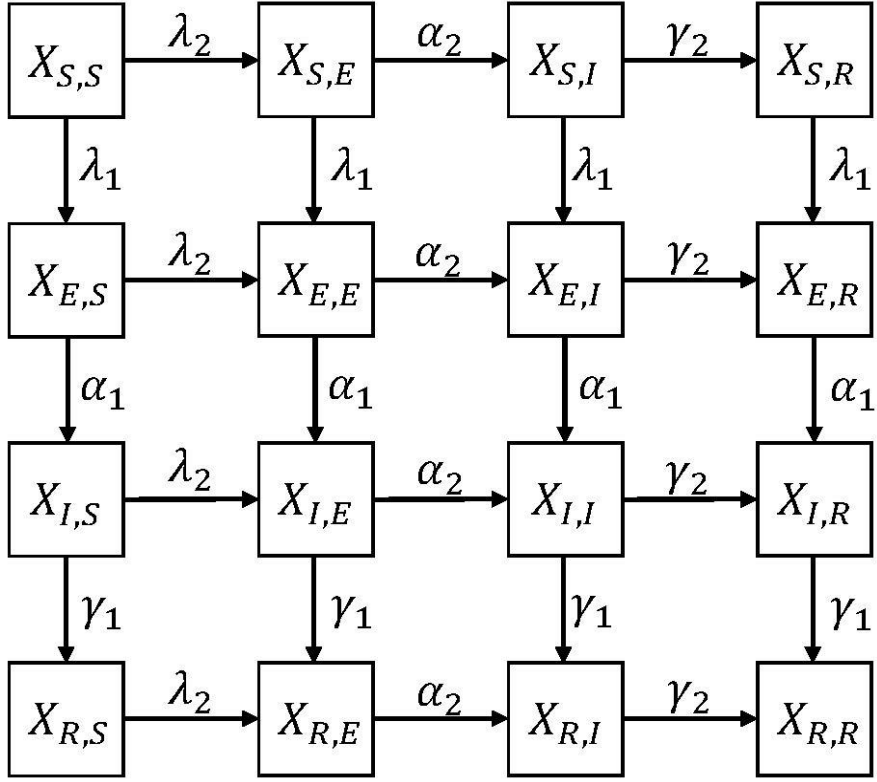

**Figure S2: Schematic of the virus-virus interactions model.** Horizontal transitions are due to virus 2 dynamics, and vertical transitions due to virus 1 dynamics.

The forces of infection ($\lambda$) and prevalences ($p$) of each virus are defined as:

$$\lambda_1(t) = R_{0,1}\gamma_1 p_1(t)$$

$$p_1(t) = X_{I,S} + X_{I,E} + X_{I,I} + X_{I,R}$$

$$\lambda_2(t) = R_{0,2}\gamma_2 p_2(t)$$

$$p_2(t) = X_{S,I} + X_{E,I} + \theta X_{I,I} + X_{R,I}$$

Here, $R_{0,1}$ and $R_{0,2}$ denote the respective basic reproductive numbers for virus 1 and virus 2. $1/\alpha_1$ and $1/\alpha_2$ are the respective incubation periods and $1/\gamma_1$ and $1/\gamma_2$ the recovery periods for the viral diseases. The parameter $\theta$ is an interaction parameter indicating the impact of infection with virus 1 on transmission of virus 2.

We consider the Influenza A - SARS-CoV-2 interacting system as an example, where virus 1 is Influenza A and virus 2 is SARS-CoV-2, so infection with influenza A affects the dynamics of SARS-CoV-2, but infection with SARS-CoV-2 has no impact on influenza A. Parameter values are detailed in Table S6. The model was run for 365 days and the peak SARS-Cov-2 incidence was calculated, for varying of influenza A basic reproduction numbers and varying transmission interaction parameter.

**Table S6. Parameters used for influenza A - SARS-CoV-2 interaction model.**

| Parameter | Meaning | Fixed values | Source |
|---|---|---|---|
| $X_{E,S}(t=0)$ | Initial fraction exposed to virus 1 | $1 \times 10^{-3}$ | Assumption |
| $X_{R,S}(t=0)$ | Initial fraction immune to virus 1 | 0.2 | Assumption |
| $R_{0,1}$ | Virus 1 basic reproductive number | 1.0–2.5 | [15] |
| $1/\alpha_1$ | Virus 1 latent period | 1 day | [16] |
| $1/\gamma_1$ | Virus 1 recovery period | 4 days | [17] |
| $X_{S,E}(t=0)$ | Initial fraction exposed to virus 2 | $1 \times 10^{-5}$ | Assumption |

| Parameter | Meaning | Fixed values | Source |
|---|---|---|---|
| $R_{0,2}$ | Virus 2 basic reproductive number | 2 | [12] |
| $1/\alpha_2$ | Virus 2 latent period | 4 days | [12] |
| $1/\gamma_2$ | Virus 2 recovery period | 5 days | [13] |
| $\theta$ | Coinfection impact on viral transmission | 0.2, 0.5, 0.8, 1.0, 1.2, 2.0, 5.0 | Assumption |